\documentclass[manuscript]{aastex}

\usepackage{graphicx}
\usepackage{epsfig}
\usepackage{amsmath}
\usepackage{amssymb}
\usepackage{color}
\usepackage{url}
\usepackage{textcomp}

\newcommand{\jastp}{J.~Atm.~Sol.~Terr.~Phys.}

\newcommand{\phiA}{\phi_{A}}
\newcommand{\phiB}{\phi_{B}}
\newcommand{\thtA}{\theta_{A}}
\newcommand{\thtB}{\theta_{B}}

\newcommand{\Rsun}{\textrm{R}_{\odot}}

\shorttitle{Kinematics of Two Eruptive Prominences}
\shortauthors{Joshi \& Srivastava}

\begin{document}

\title{Kinematics of Two Eruptive Prominences observed by EUVI/STEREO}

\author{A. D. Joshi\altaffilmark{1} and N. Srivastava\altaffilmark{1}}
\affil{Udaipur Solar Observatory, Physical Research Laboratory,
P.O.~Box 198, Badi Road, Udaipur 313001, India}
\email{janandd@prl.res.in}

%%%%%%%%%%%%%%%%%%%%%%%%%%%%%%%%%%%%%%%%%%%%%%%%%%%%%%%%%%%%%%%%%%%%%%%%%
%%% Abstract
%%%%%%%%%%%%%%%%%%%%%%%%%%%%%%%%%%%%%%%%%%%%%%%%%%%%%%%%%%%%%%%%%%%%%%%%%
\begin{abstract}
Two large northern polar crown prominences that erupted on 2010 April 13 and 2010 August 1 were analysed using images obtained from the \textit{Extreme UltraViolet Imager} on the twin \textit{Solar Terrestrial Relations Observatory} spacecraft. Several features along the prominence legs were reconstructed using a stereoscopic reconstruction technique developed by us. The three-dimensional changes exhibited by the prominences can be explained as an interplay between two different motions, namely helical twist in the prominence spine, and overall non-radial equatorward motion of the entire prominence structure. The sense of twist in both the prominences is determined from the changes in latitudes and longitudes of the reconstructed features. The prominences are observed starting from a few hours before the eruption. Increase in height before and during the eruption allowed us to study kinematics of the prominences in the two phases of eruption, the slow rise and the fast eruptive phase. A constant value of acceleration was found for each reconstructed feature in each phase, but it showed significant change from one leg to the other in both the prominences. The magnitude of acceleration during the eruptive phase is found to be commensurate with the net effect of the two motions stated above.
\end{abstract}


%%%%%%%%%%%%%%%%%%%%%%%%%%%%%%%%%%%%%%%%%%%%%%%%%%%%%%%%%%%%%%%%%%%%%%%%%
%%% New section
%%%%%%%%%%%%%%%%%%%%%%%%%%%%%%%%%%%%%%%%%%%%%%%%%%%%%%%%%%%%%%%%%%%%%%%%%
\section{Introduction}\label{S:intro}

Solar prominences are formed along the polarity inversion line, also known as the neutral line, between regions of oppositely directed photospheric magnetic fields. They are supported by means of barbs, which are appendages extending from either side of the prominence spine connecting the prominence to the chromosphere \citep{martin1998a}. Prominences or filaments almost always end their life on the Sun by means of an eruption \citep{filippovden2001}. A filament may end its lifetime as a disparition brusque \citep{raaduea1987,schmiederea2000}, in which the filament diffuses slowly and disappears. In addition to the fast-rise phase during the eruption \citep{tandbergea1980,sterlingea2007}, prominences are also reported to show a slow-rise phase prior to the actual eruption \citep{schrijver2008}, either with constant velocity \citep{sterlingmoore2005} or with constant acceleration \citep{joshisrivastava2007}. \citeauthor{sterlingmoore2004a} (\citeyear{sterlingmoore2004a,sterlingmoore2004b}) have observed constant velocity for both the phases of filament eruption, and have attempted to fit models of reconnection to the observed events. \citet{grechnevea2006} have explained the filament and coronal mass ejection (CME) eruption as a three stage process with the help of a dual-filament CME initiation model.

Various studies have shown prominences to be erupting in many different ways. During the eruption stage, prominences are known to exhibit a twist \citep{vrsnakea1991,vrsnakea1993}. This has been further observed by \citet{srivastavaea1991} and \citet{gilbertea2007}, who have explained the helical structure by means of kink instability. \citet{srivastavaambastha1998} have determined several physical parameters of a helically twisted prominence. Prominences, at times, can also erupt asymmetrically, i.e., one leg remains fixed in the lower corona, while the other leg is seen to erupt \citep{tripathiea2006}. CMEs which are known to be closely associated with eruptive prominences (\citeauthor{gopalswamyea2003} \citeyear{gopalswamyea2003} and references therein) are also shown to exhibit twisted helical structures \citep{dereea1999}.

In this study, we focus mainly on two aspects of the prominence eruption, namely the kinematics during the two rise phases, and the helical twist of prominences during the fast-eruptive phase. By twist we mean the filament axis leaving its plane and forming a loop-like structure, such as seen in the Transition Region and Coronal Explorer (TRACE) images, e.g., Figure 1 of \citet{torokkliem2005} and Figure 3 of \citet{chiforea2006}. Here, we feel it is necessary to distinguish between the twist that we study, and the twist observed in another dynamic phenomenon, known as the roll effect \citep{martin2003}. During roll effect, a prominence is seen to roll at the top giving rise to twists in mutually opposite directions in the two legs of the prominence. In our present study, we concentrate mainly on the twisting nature of the two erupting prominences, and quantify it in terms of the changes in latitude and longitude of features selected along their legs.

It should be noted that almost all of the studies, cited above, involving twist of filaments during their eruption, and the two phases of rise have been done either using ground-based data, or data from a single spacecraft. Projection effects due to a single point of view are inherent in such studies. We use observations from the identical \textit{Extreme UltraViolet Imager} (EUVI) \citep{howardea2008} instruments on board the twin \textit{Solar TErrestrial RElations Observatory} (STEREO) spacecraft \citep{kaiserea2008} for a three-dimensional study. Researchers have used data from the STEREO spacecraft to study various aspect of prominence dynamics. \citet{gissotea2008} and \citet{liewerea2009} have carried out stereoscopic studies to obtain true coordinates and hence the true velocity of the prominence on 2007 May 19. \citet{bemporad2009} and \citet{liea2010} have reconstructed several features of the prominence during its eruption to study the prominence shape as a whole. On the other hand, \citet{thompson2010} has observed rotation of the prominence about its direction of eruption. \citet{panasencoea2010} have studied the rolling motion of three prominences and the associated CMEs from stereoscopic reconstruction applied to STEREO observations.

In the present study, two prominence eruption events on 2010 April 13 and 2010 August 1 are analysed. Both the prominences are high latitude northern hemisphere prominences which erupted over a period of a few hours. The time-lapse movies of the events show highly twisted prominence spines during the eruptions. We have used EUVI 304 \AA~ images to measure the evolution of prominences in physical coordinates. Several features along the prominence body in images from EUVI Ahead (A) and Behind (B) were selected and using stereoscopic technique developed by us, the shapes of the prominences at several instants of time were obtained. 

We have developed a new stereoscopic reconstruction technique for the images obtained from the STEREO spacecraft. We initially work with the heliocentric Earth ecliptic (HEE) coordinate system, and later convert to the more common heliographic system. The technique involves rotating the HEE system separately for STEREO A and B, such that one of the axes of the HEE system lies along the line-of-sight of each spacecraft. The plane perpendicular to this axis is therefore the image plane, i.e., the plane of sky for the concerned spacecraft. We then determine the physical coordinates of the feature in the HEE system by solving the corresponding rotation matrices. It should be noted, that since we impose no conditions on the position of the feature to be reconstructed, this technique can be applied equally well to EUVI and coronagraph images, viz, COR1 and COR2, as long as the condition for affine geometry is valid. Details of the reconstruction technique are discussed in the Appendix.

%%%%%%%%%%%%%%%%%%%%%%%%%%%%%%%%%%%%%%%%%%%%%%%%%%%%%%%%%%%%%%%%%%%%%%%%%
%%% New section
%%%%%%%%%%%%%%%%%%%%%%%%%%%%%%%%%%%%%%%%%%%%%%%%%%%%%%%%%%%%%%%%%%%%%%%%%
\section{Data Analysis}\label{S:data}

We have used 304~\AA~images from \textit{EUVI/STEREO} to reconstruct shape of the prominences on 2010 April 13 and 2010 August 1. Both the prominences were located in the northern hemisphere. In spite of the large separation between the two spacecraft, ($139^{\circ}$ on 2010 April 13 and $149^{\circ}$ on 2010 August 1), the elevated heights of the two prominences made the three-dimensional reconstruction possible. To identify a feature unambiguously in both the images, and then track it correctly in subsequent images, we have extensively relied upon time-lapse movies of the events. The movies guided us to safely neglect features that did not persist for the entire duration of the eruption, and consider only those that could be identified and tracked in all of the images. However, since both the prominences are located very close to the solar limb as seen from each spacecraft, even such a time-lapse movie is not enough to reveal the sense of twist in the prominence. Hence we use the stereoscopic reconstruction technique.

\begin{figure}[!hbp]
\centering
\includegraphics[width=0.30\textwidth,clip=]{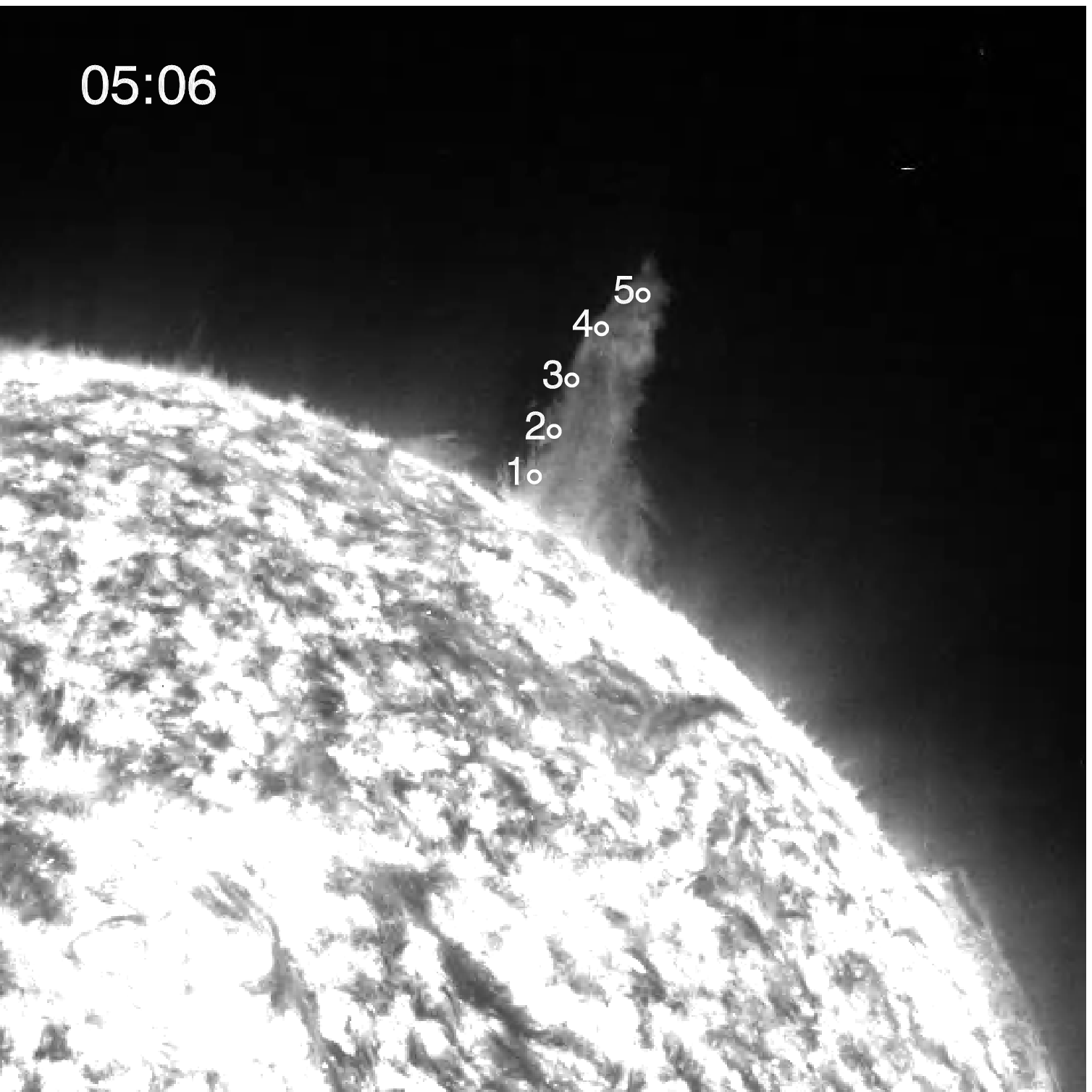}
\includegraphics[width=0.30\textwidth,clip=]{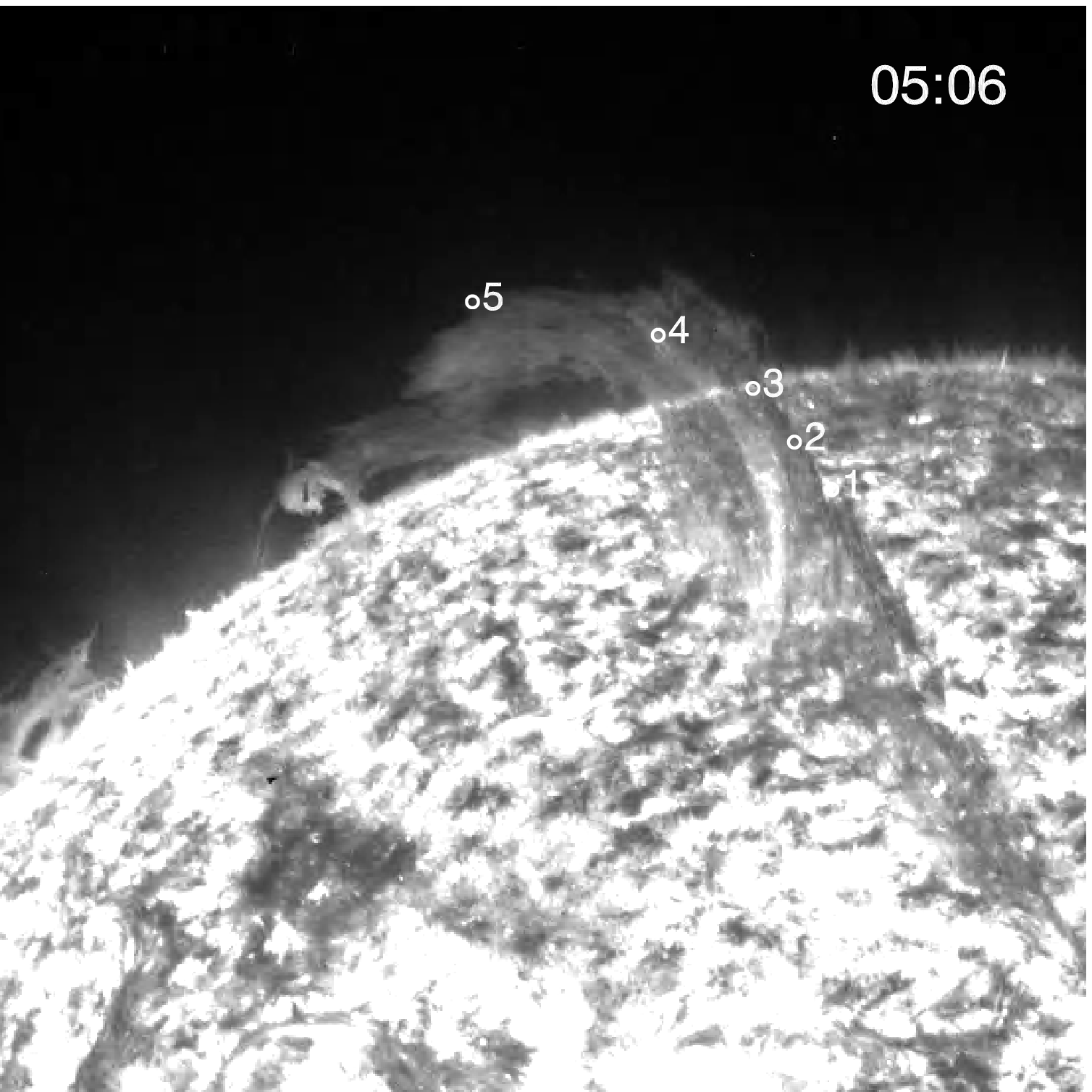}
\\
\includegraphics[width=0.30\textwidth,clip=]{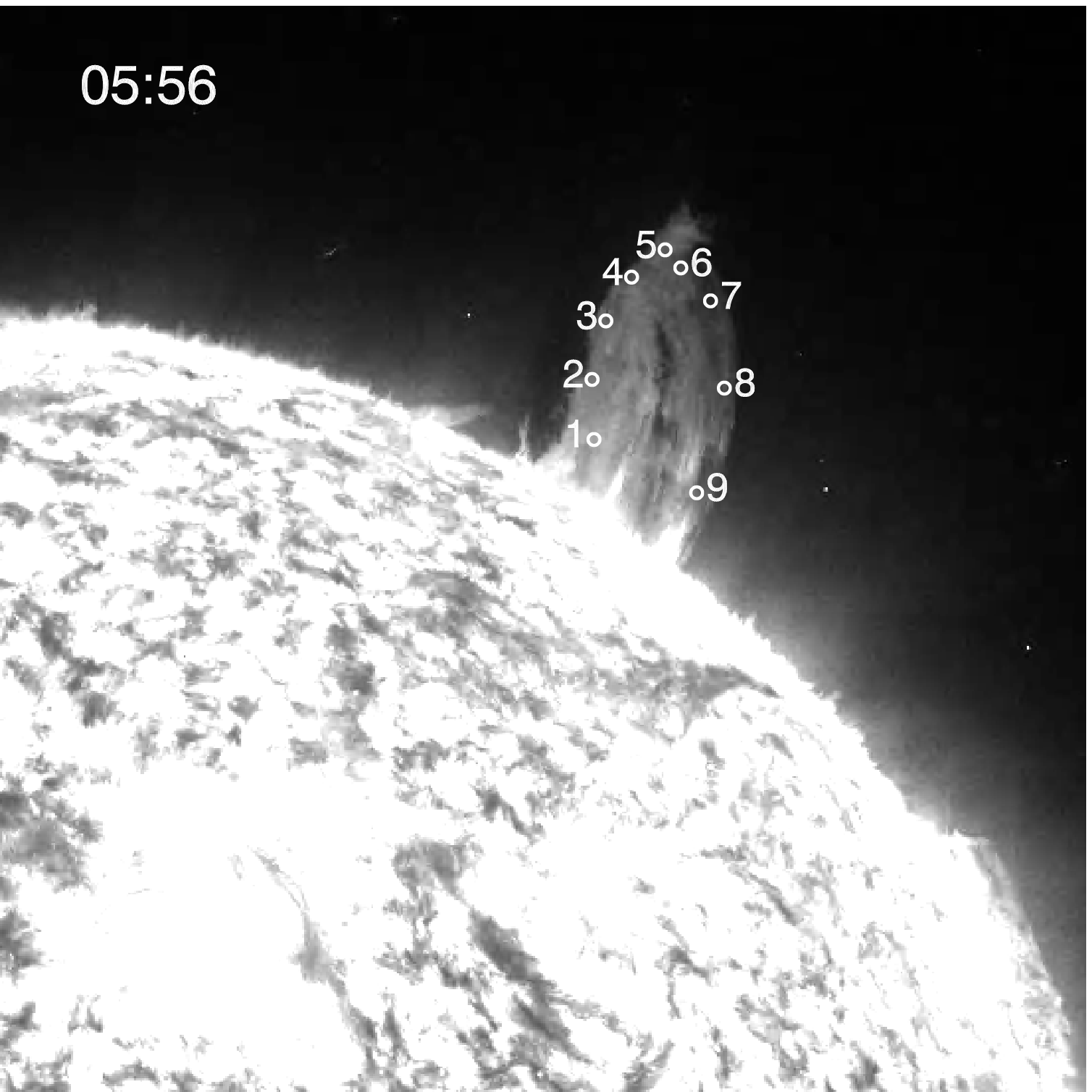}
\includegraphics[width=0.30\textwidth,clip=]{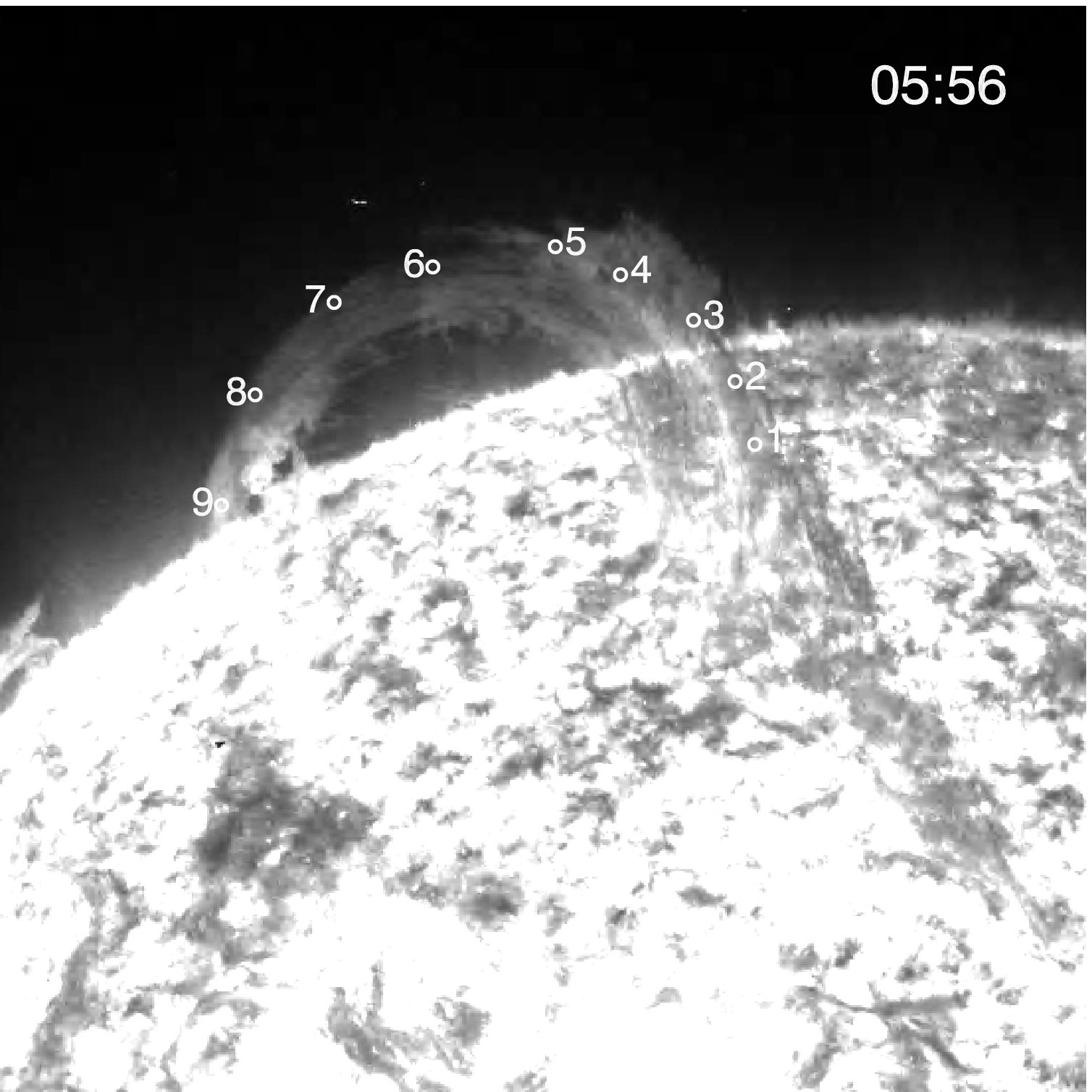}
\\
\includegraphics[width=0.30\textwidth,clip=]{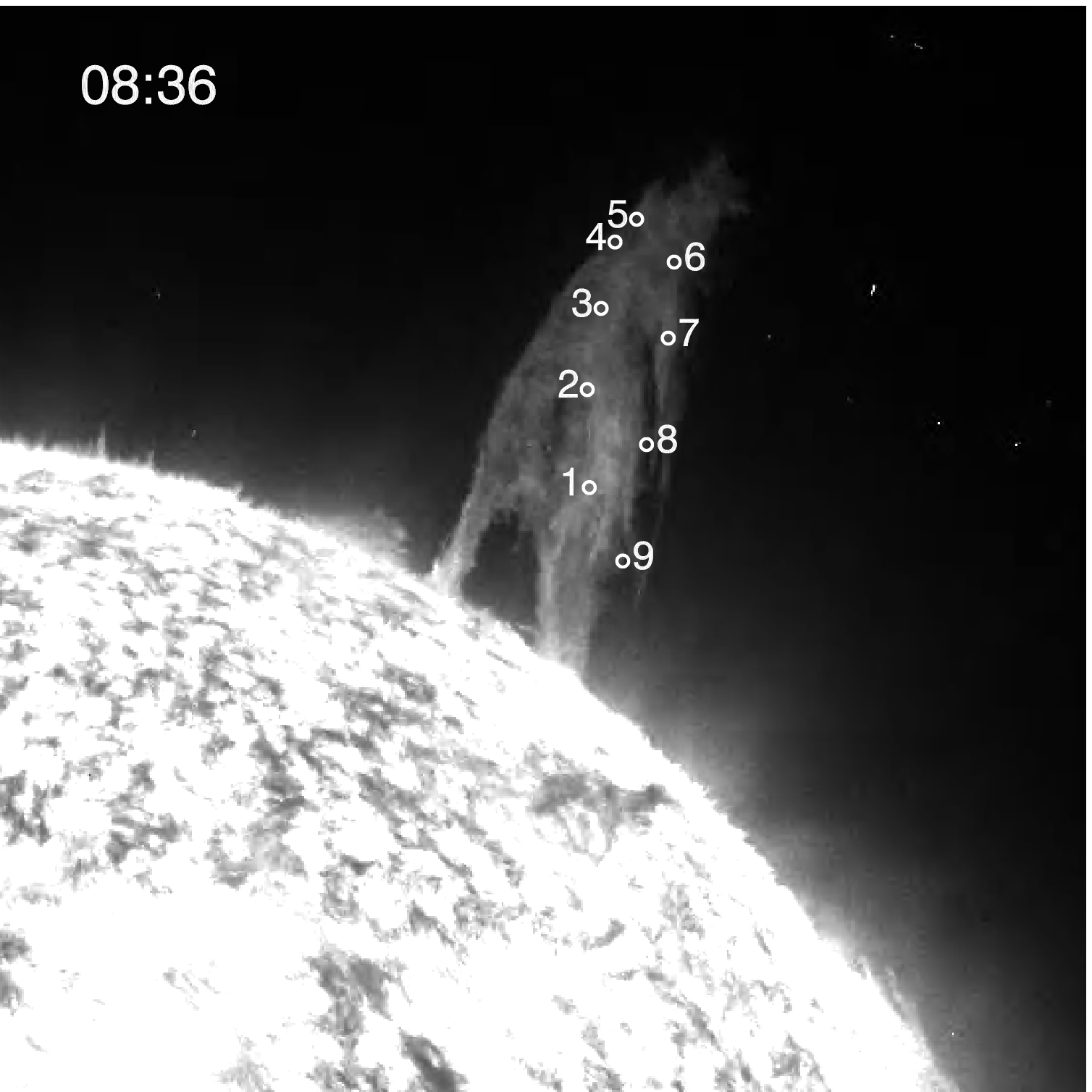}
\includegraphics[width=0.30\textwidth,clip=]{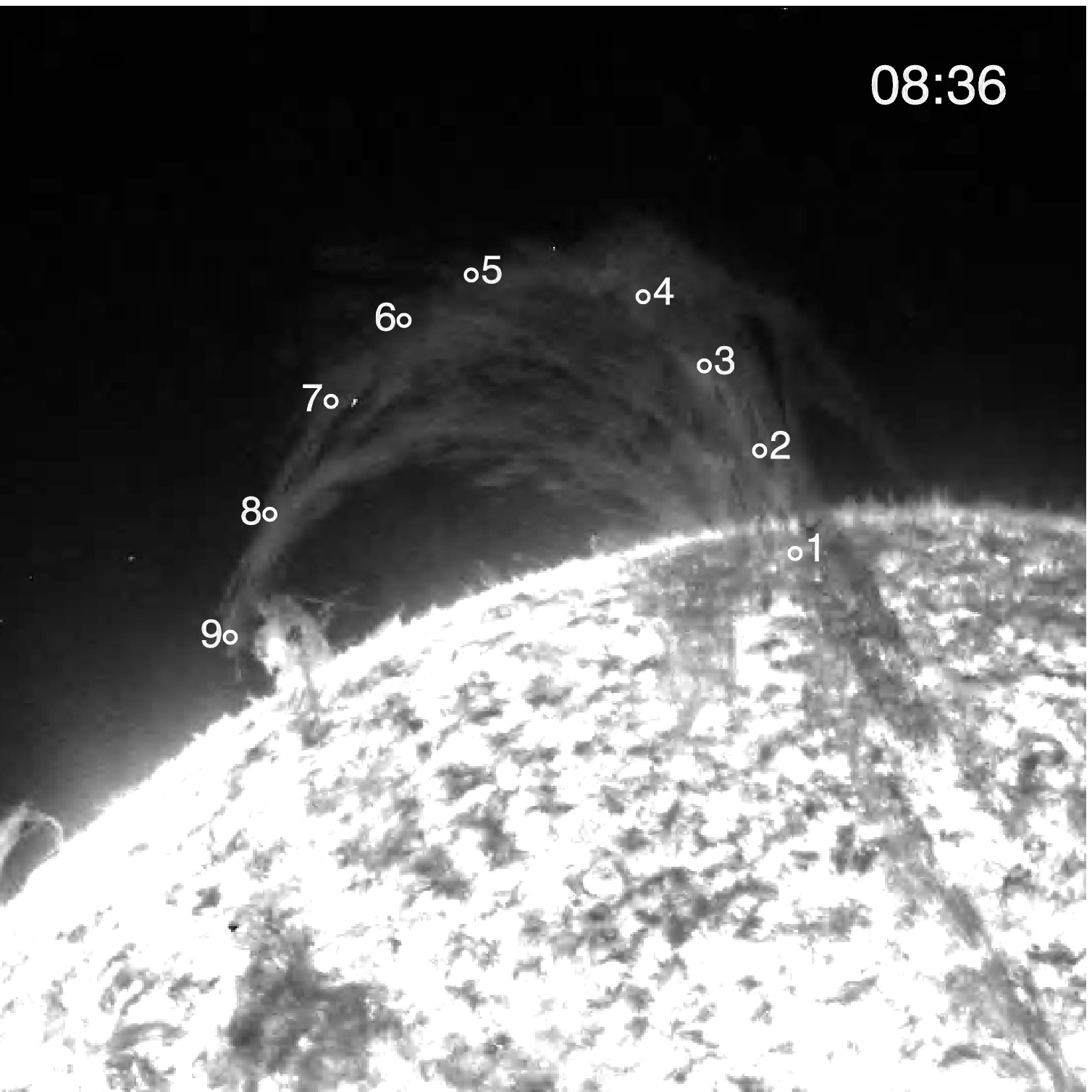}
\\
\includegraphics[width=0.30\textwidth,clip=]{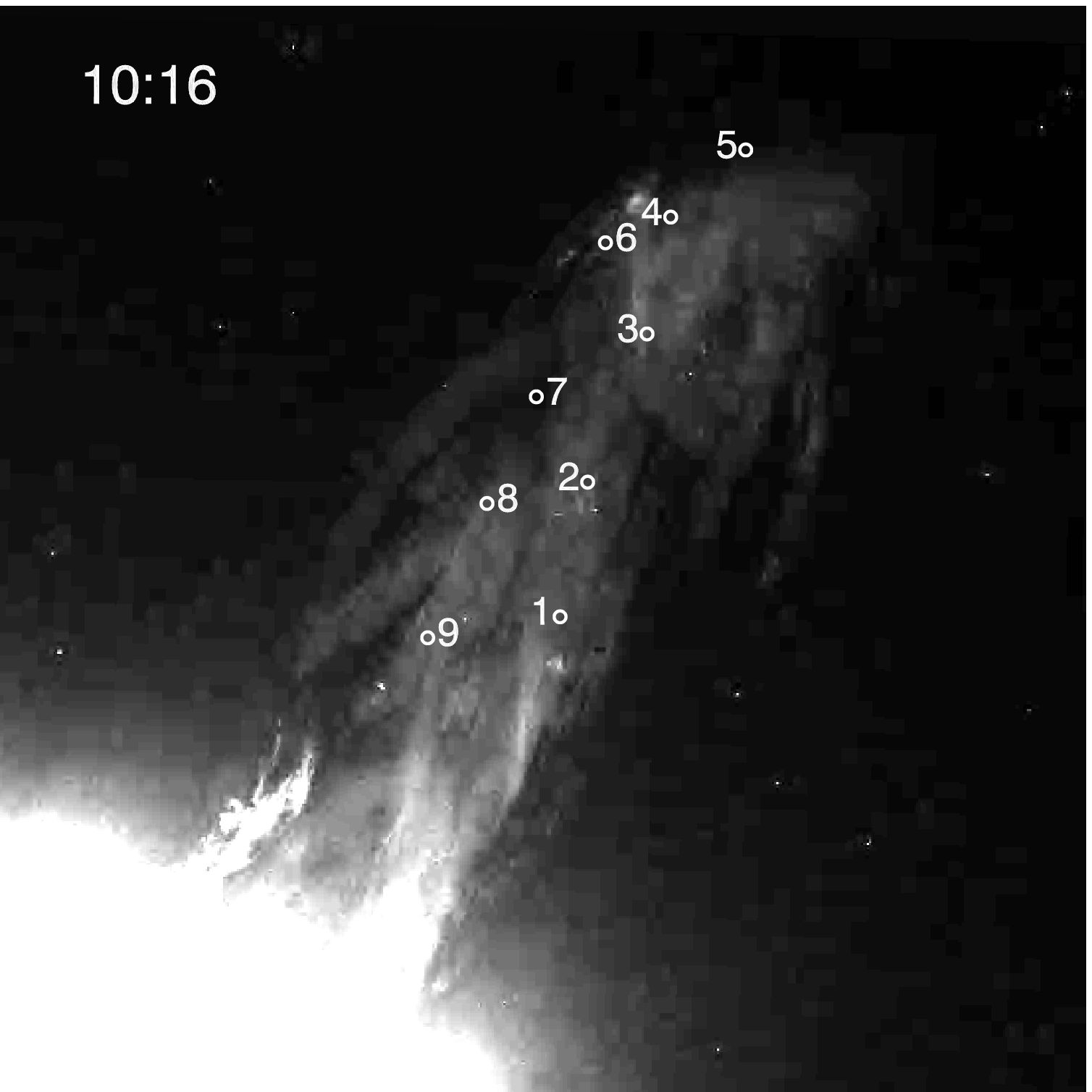}
\includegraphics[width=0.30\textwidth,clip=]{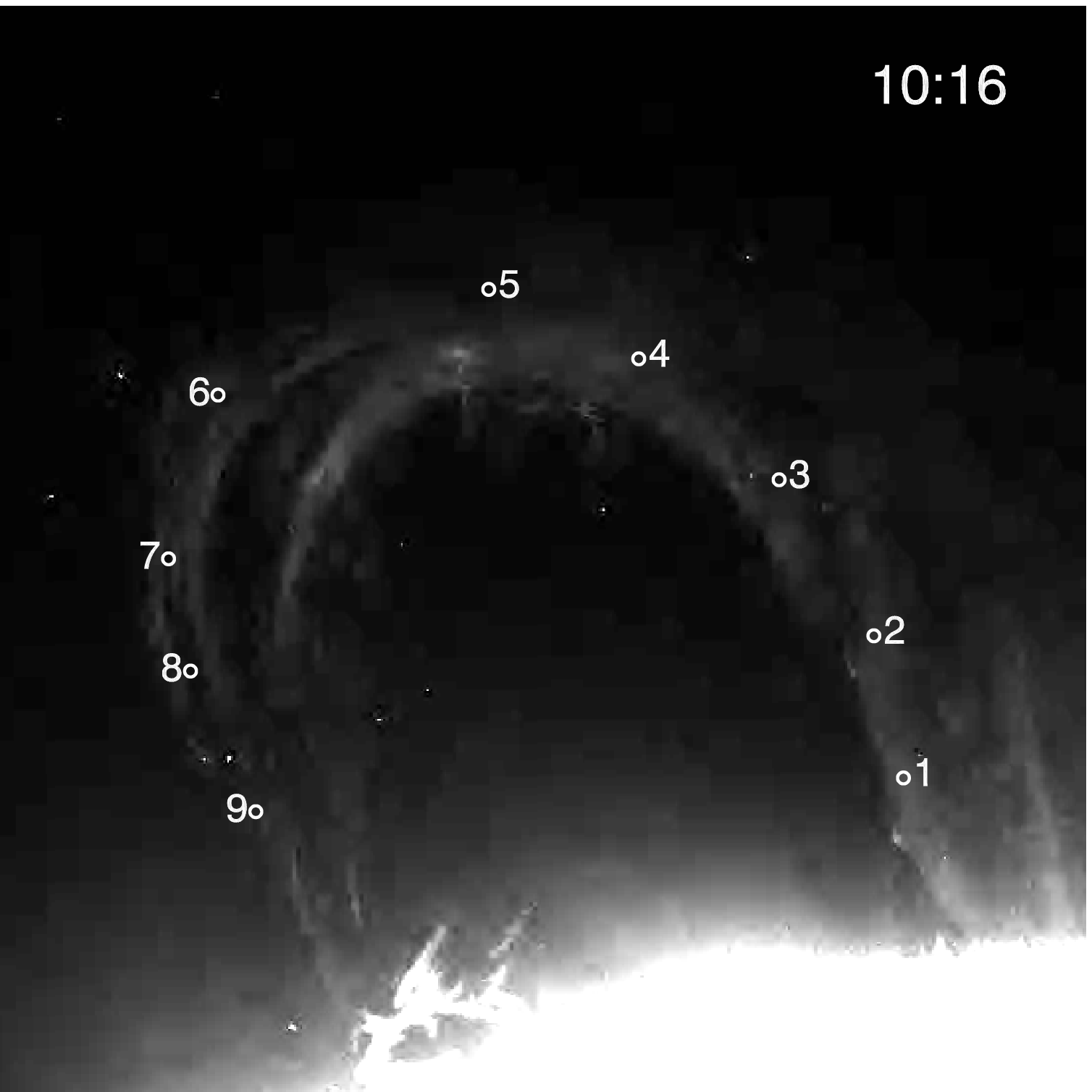}
\\
\caption{Erupting prominence on 2010 April 13 seen in 304~\AA~images from EUVI B (left) and A (right) on board the twin STEREO spacecraft. Observation times in UT are shown for each image. The features used for reconstruction are marked and numbered along the prominence. Leg L2 was visible in EUVI B only from \mbox{05:36 UT} onwards, hence numbers 6 to 9 are not shown in the image at \mbox{05:06 UT}.}\label{F:img13apr}
\end{figure}

\subsection{Event of 2010 April 13}

This is a high-latitude northern polar crown prominence. Images from EUVI B (left column) and A (right column) at different instants of time are shown in Figure~\ref{F:img13apr}. We name the prominence leg on the right hand side in EUVI A image in Figure~\ref{F:img13apr} as L1 and the one on the left (in EUVI A image) as L2. At \mbox{05:06 UT}, we see that the prominence is oriented in such a direction, that leg L2 is not visible from EUVI B. In EUVI A however, the prominence appears almost side-on, giving us a complete view of its evolution. At \mbox{05:56 UT} L2 starts to show up in EUVI B images. Mass flows in L2 are observed during this time in both the images. 

We chose five features along the leg L1 of this prominence to be reconstructed numbered 1 to 5 from bottom of the leg up to the top of the spine. In addition, once leg L2 was visible in both the images, from \mbox{05:36 UT} onwards, we identified four features along it, numbered 6 to 9 starting from top of the spine and reaching bottom of L2, and followed those too. The features are marked and numbered on the images in Figure~\ref{F:img13apr}. As the eruption progressed, the prominence became more twisted, and also grew fainter as it rose in height. The features selected for reconstruction were followed carefully until it was no longer possible to identify them unambiguously.

\subsection{Event of 2010 August 1}

\begin{figure}[!hbp]
\centering
\includegraphics[width=0.30\textwidth,clip=]{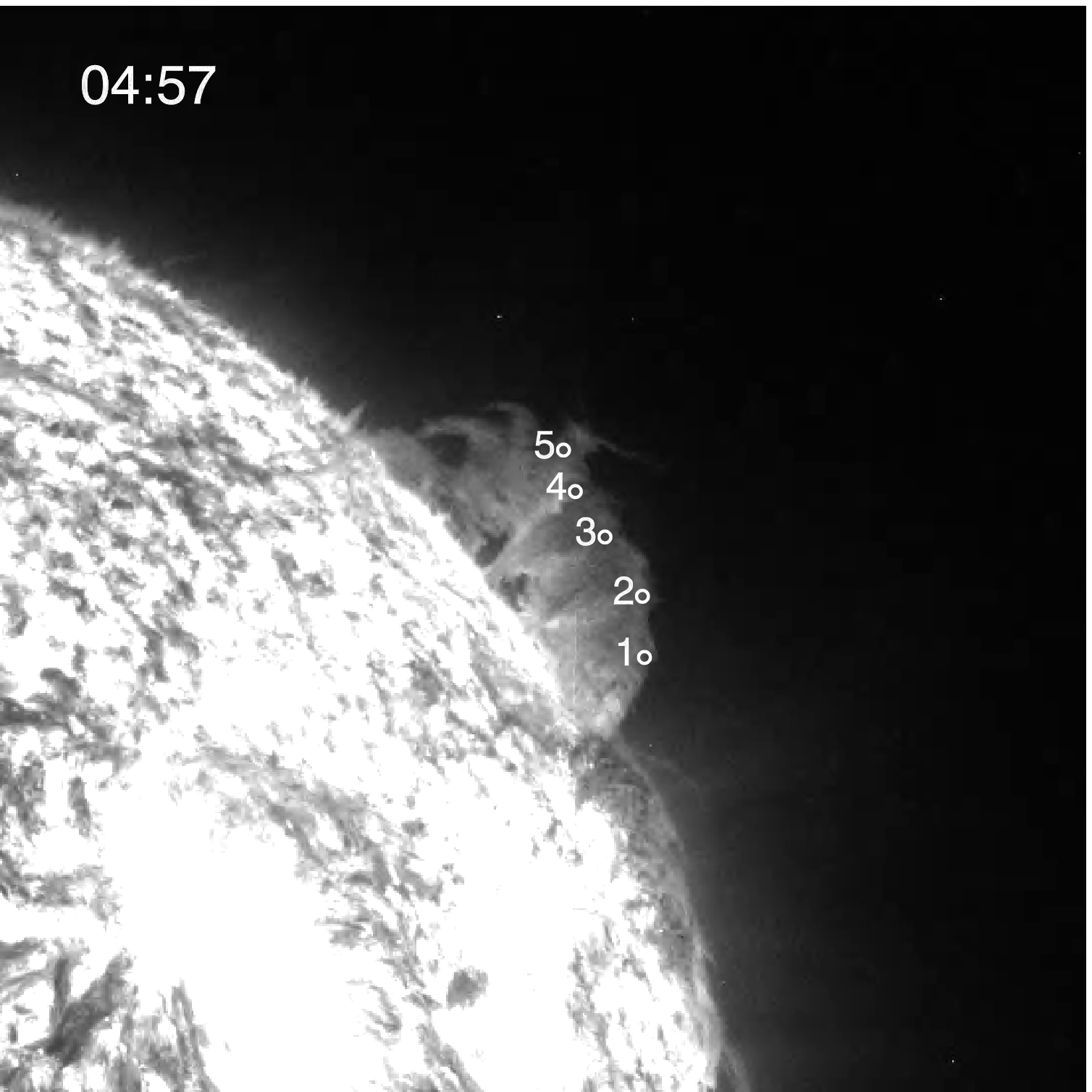}
\includegraphics[width=0.30\textwidth,clip=]{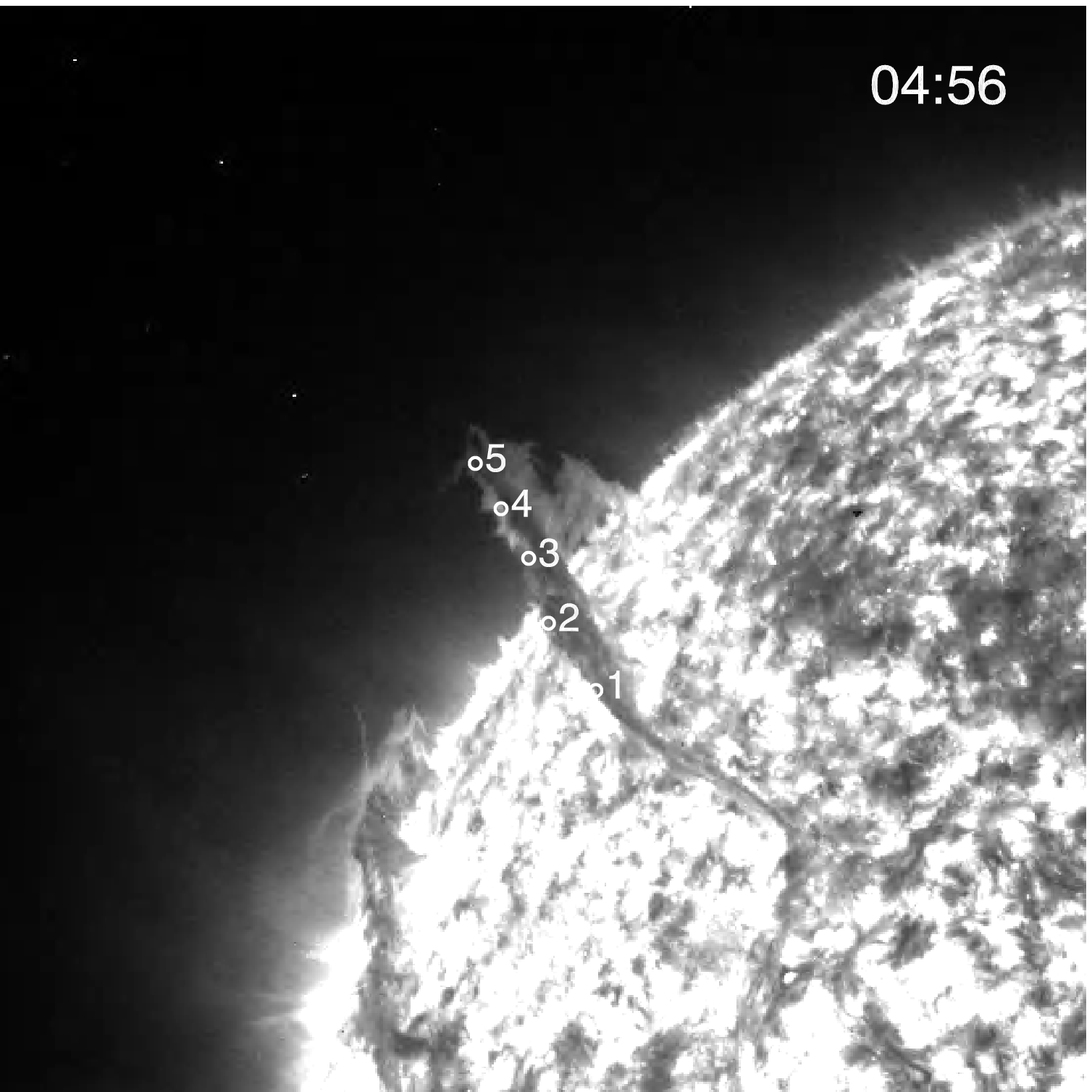}
\\
\includegraphics[width=0.30\textwidth,clip=]{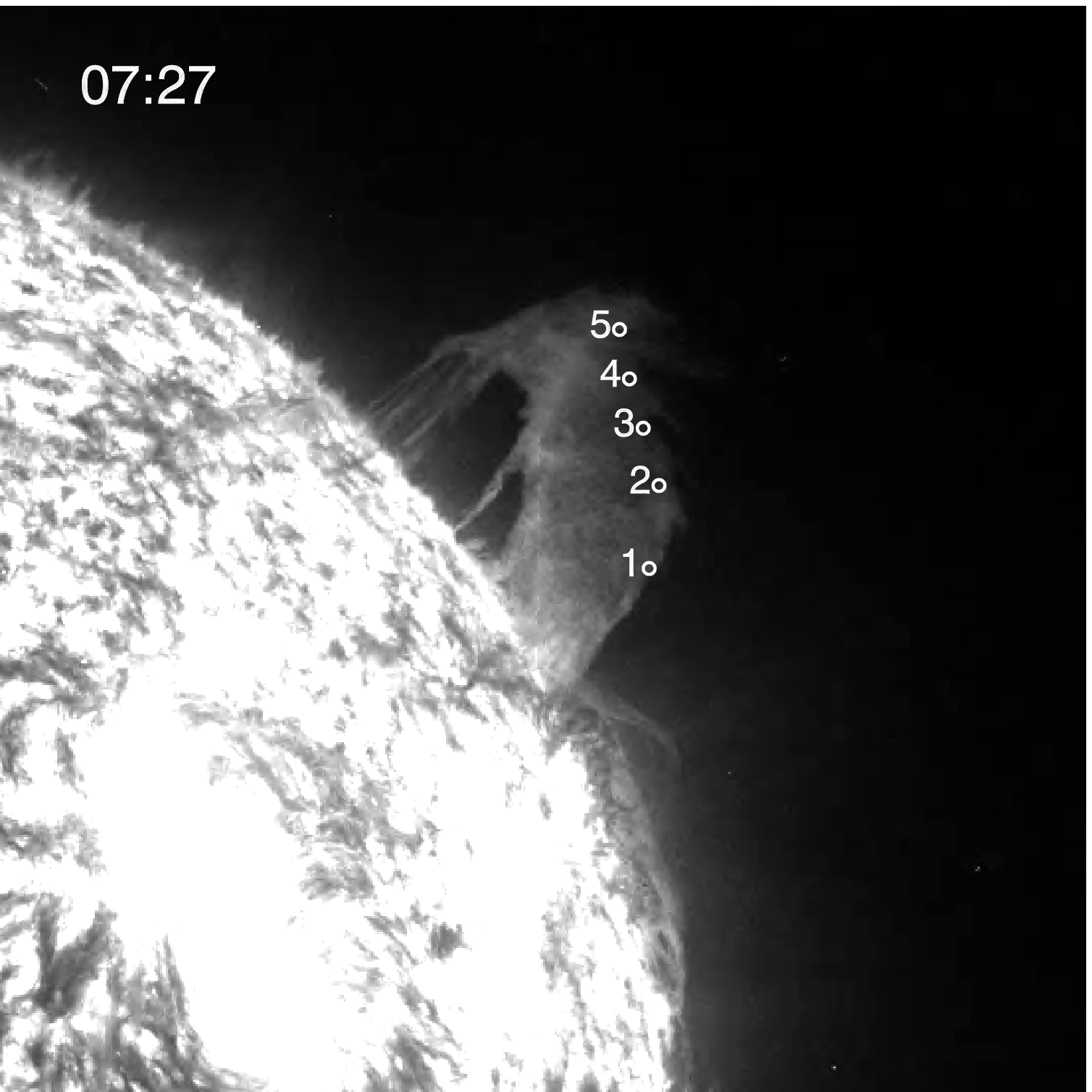}
\includegraphics[width=0.30\textwidth,clip=]{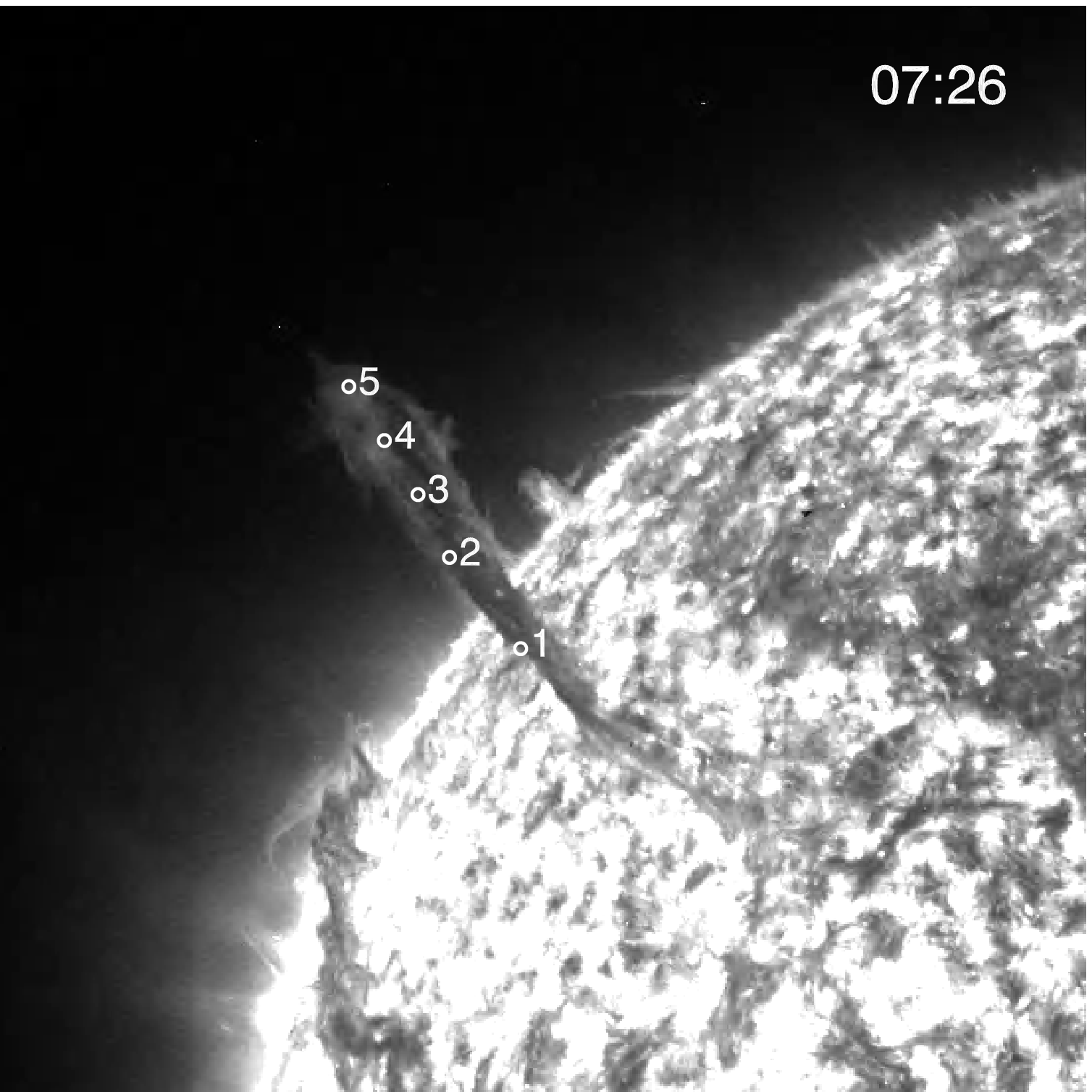}
\\
\includegraphics[width=0.30\textwidth,clip=]{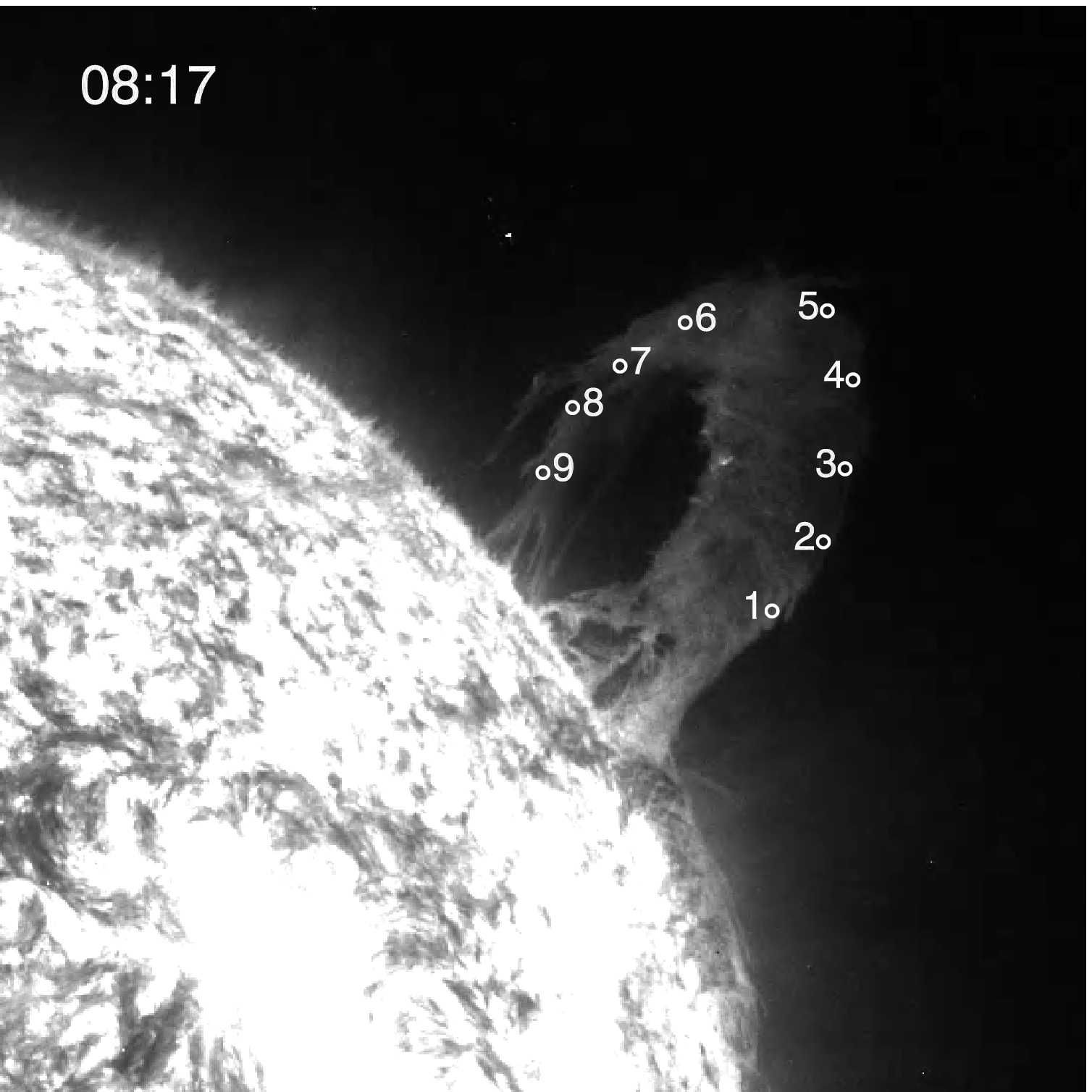}
\includegraphics[width=0.30\textwidth,clip=]{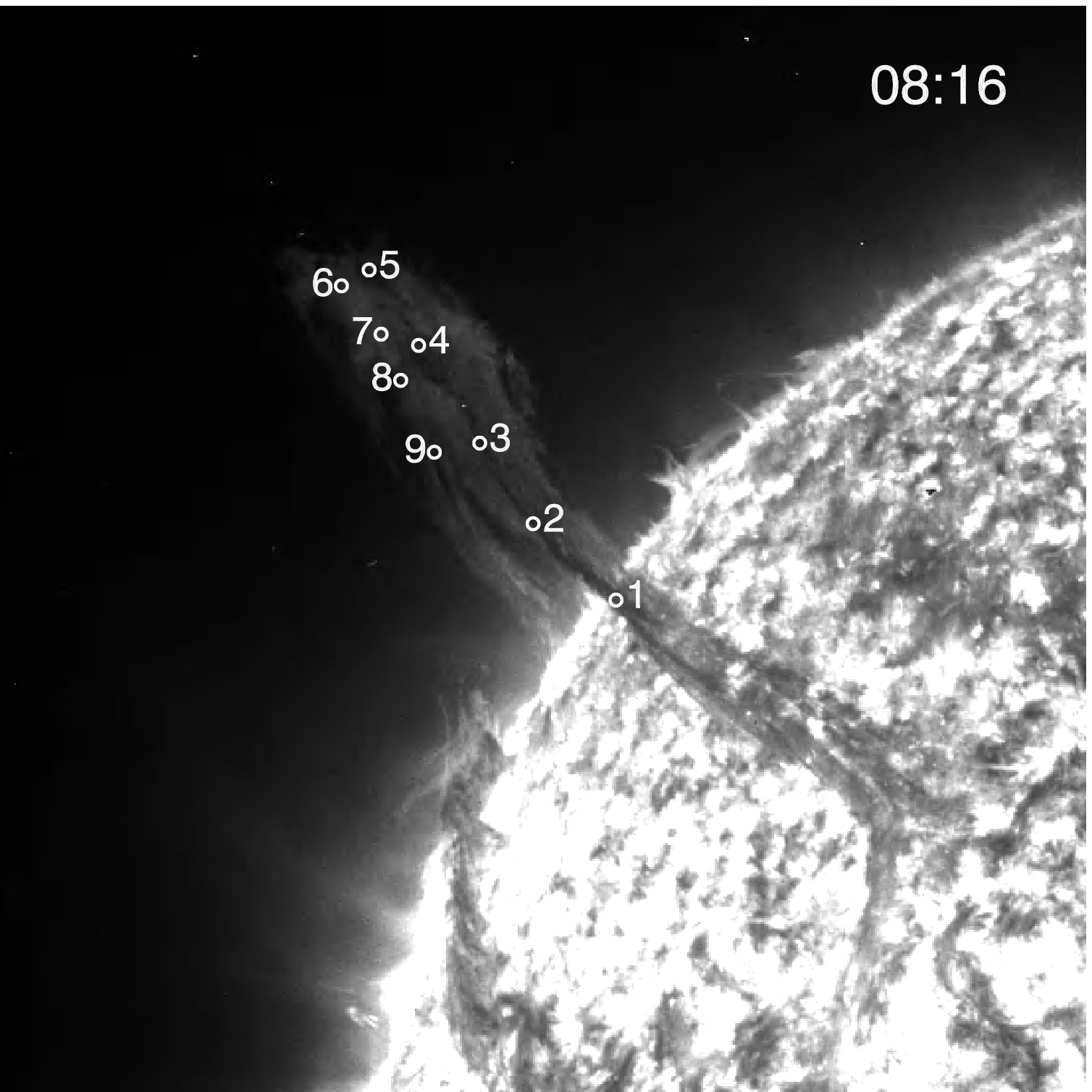}
\\
\includegraphics[width=0.30\textwidth,clip=]{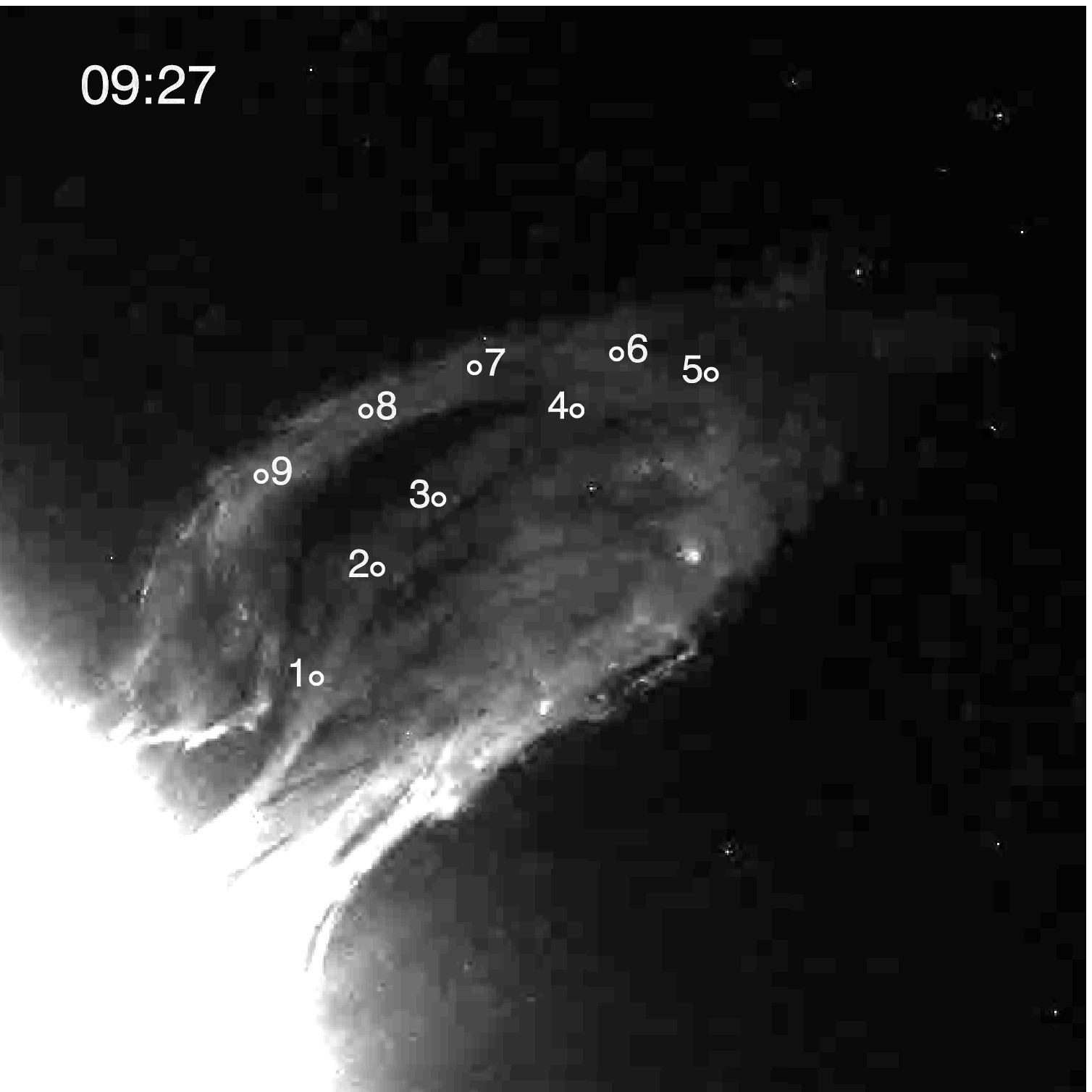}
\includegraphics[width=0.30\textwidth,clip=]{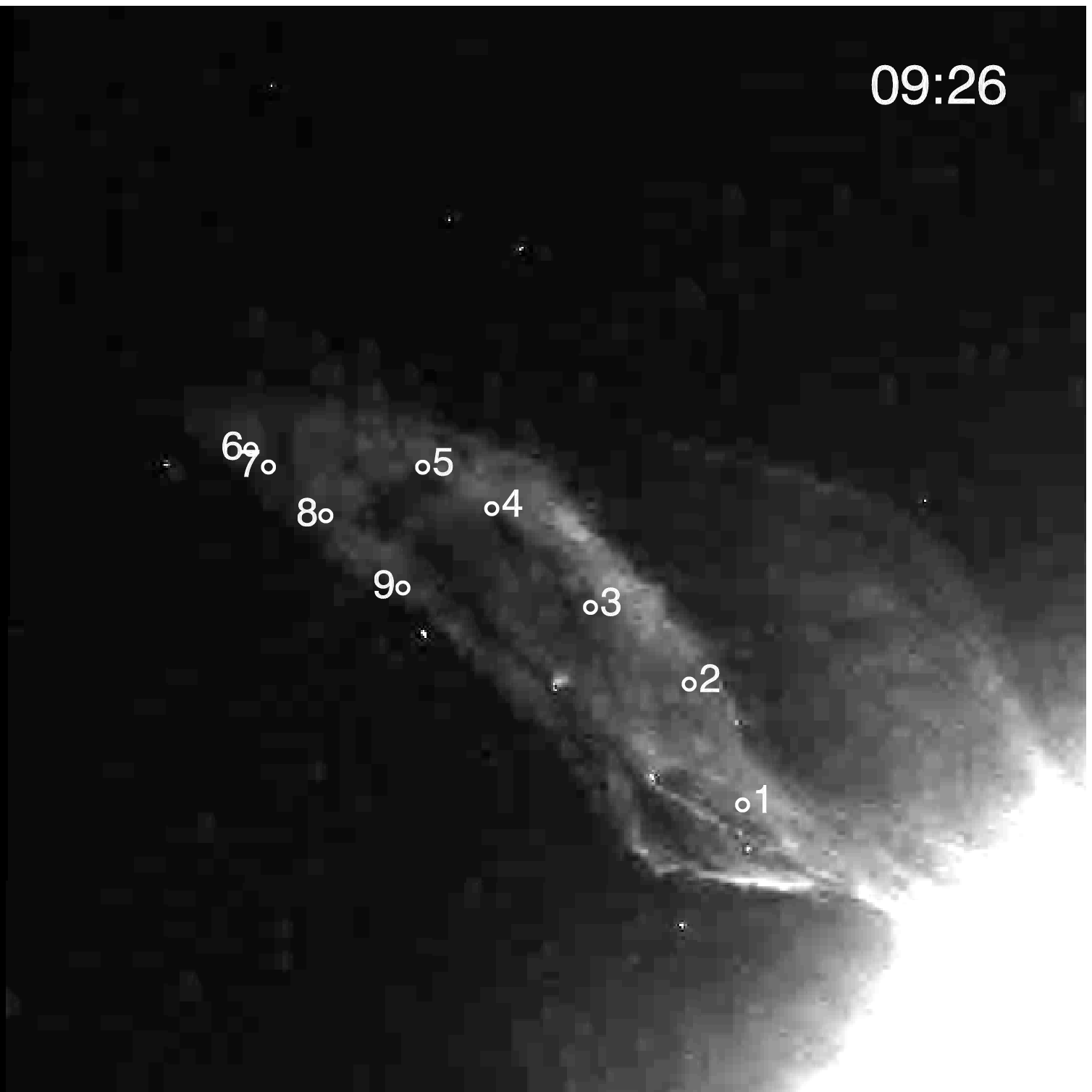}
\\
\caption{Erupting prominence on 2010 August 1 seen in 304~\AA~images from EUVI B (left) and A (right) on board the twin STEREO spacecraft. Observation times in UT are shown for each image. The features used for reconstruction are marked and numbered along the prominence. Leg L2 was visible in EUVI A only from \mbox{07:36 UT} onwards, hence numbers 6 to 9 are not shown in the images at \mbox{04:56 UT} and \mbox{07:26 UT}. }\label{F:img01aug}
\end{figure}

This too is a northern polar crown prominence. 304~\AA~images of the prominence at different instants of time from EUVI B (left column) and A (right column) can be seen in Figure~\ref{F:img01aug}. We name the prominence leg seen on the right hand side in EUVI B image Figure~\ref{F:img13apr} as L1, and the one on the left as L2. While the prominence is seen as a hedgerow in EUVI B image at \mbox{04:57 UT}, we can see only its spine in the corresponding EUVI A image. At \mbox{07:27 UT}, as the prominence starts to rise, we can see an arch in EUVI B image. By this time, in EUVI A, the leg L2 of the prominence is not visible because the line-of-sight of the spacecraft is along the prominence spine. At around \mbox{08:16 UT}, the rising prominence starts to twist, and L2 can be seen in EUVI A as well. Further, at \mbox{09:26 UT} we can clearly see the twist in the prominence legs in 304~\AA~image from EUVI A.

Leg L1 of the prominence can be seen clearly in all the images from EUVI B and A, hence we chose five features along this leg for reconstruction, numbered 1 to 5 from bottom of the leg up to the top of the spine. Once the eruption starts and the prominence undergoes twisting motion, leg L2 becomes visible from \mbox{07:36 UT} onwards. We consider four features along this leg for reconstruction, numbered 6 to 9 starting from top of the spine and reaching bottom of L2. This gives us a complete picture of the true shape of the prominence.

\subsection{Three-Dimensional Reconstruction}

\citet{jacksonfroehling1995} were among the earliest ones to carry out stereoscopy for solar observations. They used data from the Solwind and Helios spacecraft to reconstruct a CME. Over the past few years, the coordinated simultaneous observations from STEREO spacecraft have greatly aided reconstruction of solar features. \citet{aschwandenea2008} employed reconstruction technique on STEREO/EUVI images to obtain three dimensional coordinates of coronal loops. \citet{thompson2009} has developed a graphical user interface in the routine {\sf{scc\_measure}} in the Solar SoftWare (SSW) library of Interactive Data Language (IDL$^{\rm TM}$), which uses tie-pointing method \citep{inhester2006} for reconstructing coronal features. \citet{mierlaea2008} have developed a 3D height-time reconstruction technique for CME features based on their height-time measurements. \citet{howardtappin2008} have also developed a triangulation technique for reconstructing CMEs using combined observations from SoHO/LASCO \citep{bruecknerea1995} and STEREO A and B. \citeauthor{liewerea2010a} (\citeyear{liewerea2010b,liewerea2010a}) have demonstrated the use of tie-pointing and triangulation technique to obtain true coordinates from EUVI and coronagraph images, on board the STEREO spacecraft. \citet{thernisienea2009} have employed forward modelling based on the graduated cylindrical shell model of CMEs to fit the best model to an observed CME, from which its evolution in three dimensions can be determined.

In the present study, we have developed a reconstruction technique based on triangulation. While developing our technique we impose no restriction on the location of a feature to be reconstructed. It is thus applicable to both features on the disc, and those in the corona. Although not demonstrated in this study, this technique can be applied to a filament eruption on the solar disc and the associated coronal mass ejection.

In this technique, we assume that the STEREO mission plane (plane containing the two STEREO spacecraft) and the ecliptic plane are the same. This is a valid assumption, since angle between the two planes never exceeds $0^{\circ}.5$, which can be neglected. We also assume an affine geometry, wherein the spacecraft are assumed to be at an infinite distance from the Sun. This too is valid, since the distances we consider in this study are up to 2~$\Rsun$ from the Sun's centre, while distances of the STEREO spacecraft from the Sun are greater than 200~$\Rsun$. Let $\phiA$ and $\thtA$ be longitude and latitude of STEREO A, and $\phiB$ and $\thtB$ be longitude and latitude of STEREO B in the HEE system, and $P(x,y,z)$ be coordinates of the point to be reconstructed. If $(x''_{A},y''_{A})$ are coordinates of point $P$ in image from STEREO A in physical units, and $(x''_{B},y''_{B})$ are coordinates of $P$ from STEREO B in physical units, then we have:

\begin{align}
x &= \frac{x''_{B}\sin{\phiA}\,-\,x''_{A}\sin{\phiB}}
	 {\sin{(\phiA-\phiB)}} \label{Q:heex}\\
y &= \frac{y''_{A}}{\cos{\thtA}}\, + 
	 \,\Big(\frac{x''_{B}}{\sin(\phiA-\phiB)} -
	 \frac{x''_{A}}{\tan{(\phiA-\phiB)}}\Big)\tan{\thtA}\\
z &= \frac{x''_{B}\cos{\phiA}\,-\,x''_{A}\cos{\phiB}}
	 {\sin{(\phiA-\phiB)}}
\label{Q:xyz}
\end{align}
The derivation of the above solution is given in detail in the Appendix. On taking into account a human error of 3 pixels in correctly identifying a  feature from a pair of EUVI images, the corresponding error in its true height, longitude, and latitude comes out to be $0.02~\Rsun$, $2^{\circ}$ and $0^{\circ}.5$ respectively.

%%%%%%%%%%%%%%%%%%%%%%%%%%%%%%%%%%%%%%%%%%%%%%%%%%%%%%%%%%%%%%%%%%%%%%%%%
%%% New section
%%%%%%%%%%%%%%%%%%%%%%%%%%%%%%%%%%%%%%%%%%%%%%%%%%%%%%%%%%%%%%%%%%%%%%%%%
\section{Results and Discussion}\label{S:resul}

\begin{figure}[!hbp]
\centering
\includegraphics[width=0.60\textwidth,clip=]{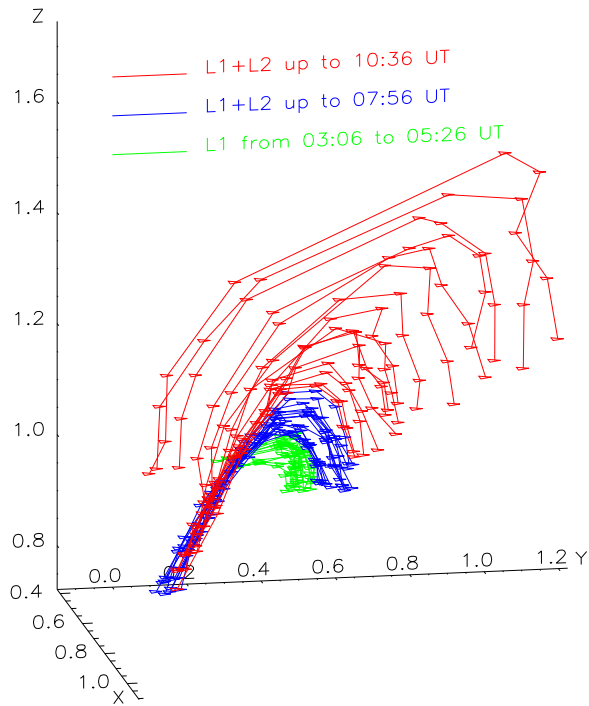}
\caption{Evolution of the erupting prominence on 2010 April 13 as seen in three dimensions in heliographic coordinate system. The position of the prominence determined by joining all the reconstructed points is shown at different instants of time, as marked on the plot.
% Cyan colour shows positions of five features in leg L1 from 03:06 to 06:16 UT; red colour shows positions of all the nine features up to 07:56 UT; blue colour shows positions of all the nine features up to 10:56 UT. 
The coordinate system is centred on the Sun, with the Z axis along the solar rotation axis, and the X axis pointing towards the Earth. All the axes are in units of $\Rsun$.}\label{F:3D13apr}
\end{figure}

\begin{figure}[!hbp]
\centering
\includegraphics[width=0.60\textwidth,clip=]{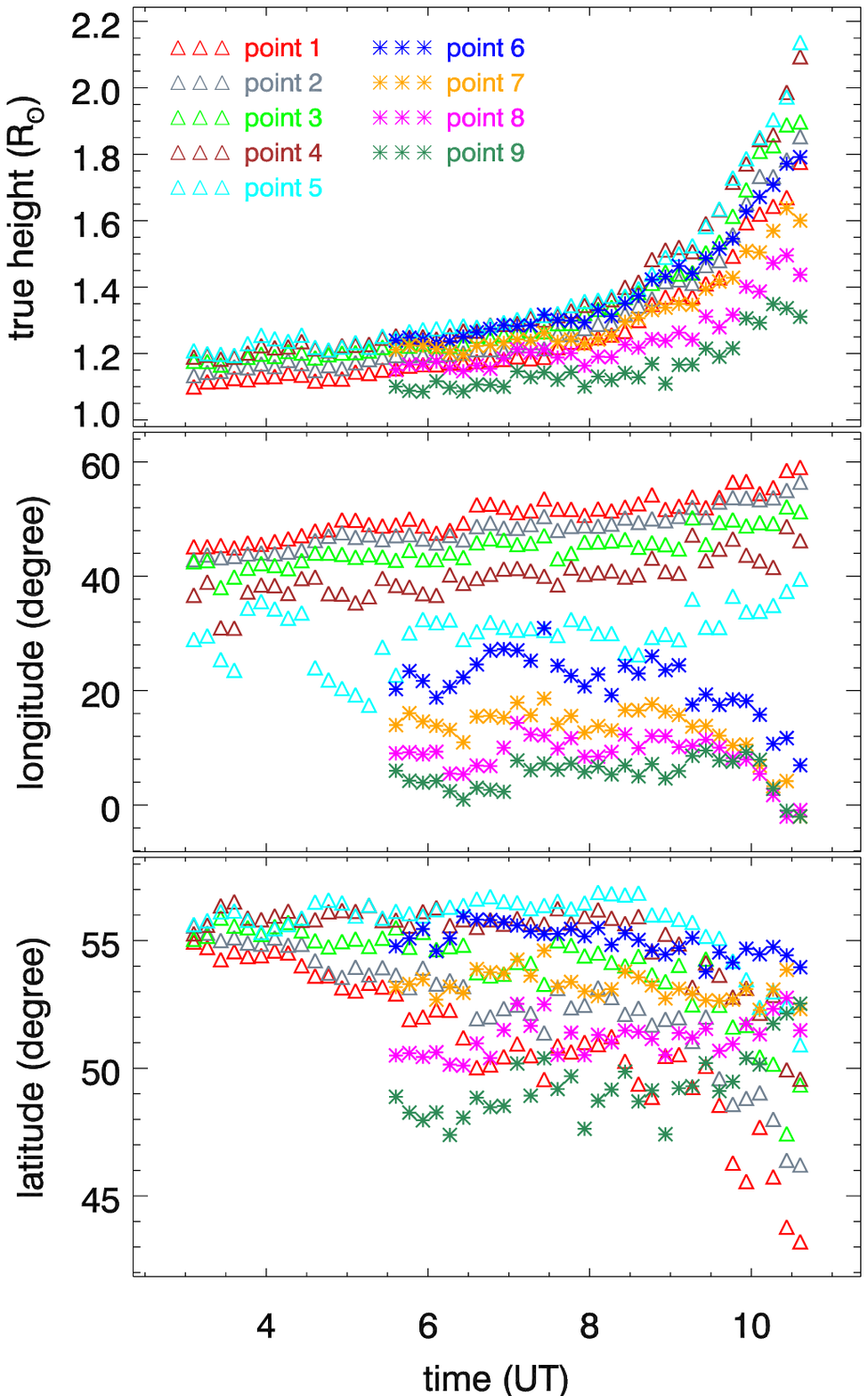}
\caption{The heliographic coordinates of different features of the prominence on 2010 April 13. Features 1 to 5 are from leg L1 to the top of spine, while features 6 to 9 are from spine to the lowest feature in leg L2 of the prominence.}\label{F:mb13apr}
\end{figure}

\subsection{2010 April 13 Prominence}

This prominence is seen nearly side-on in 304~\AA~images from EUVI A, while the line-of-sight is along the spine for EUVI B. The prominence leg seen on the right in EUVI A image at \mbox{05:56 UT} in Figure~\ref{F:img13apr} is named L1, and the other is named as L2. We chose five features in L1 and four in L2 for reconstruction. Leg L2 which was obstructed from view by L1, became visible from 05:36 UT onwards. Hence, the four features along this leg could be observed only from this time onwards. We have employed two means through which the true shape of the prominence can be visualised. Figure~\ref{F:3D13apr} shows the prominence evolution in three dimensions in heliographic coordinates. All the points reconstructed for a given time are shown to be connected by straight lines. Since, cadence for the entire duration is constant at 10 minutes, we see from this figure that the prominence rises slowly for over five hours (several closely spaced lines) before erupting rapidly in two hours (widely-spaced lines). On the other hand, Figure~\ref{F:mb13apr} shows the true height along with the longitude and latitude in heliographic coordinates of each of the nine features at different instants of time, where features 1 to 5 belong to L1, while features 6 to 9 belong to L2.

From Figure~\ref{F:mb13apr}, we can study in detail the evolution of the whole prominence structure. All the features are shown in different colours. In addition, to distinguish between features selected along the two legs, we have used different symbols. Triangles and asterisks are used for features in L1 and L2, respectively. Feature 1 is the lowest one in L1, and we can see that clearly in the plot for true height. As we move up L1 to the top of the spine, up to feature 5, the heights consistently increase throughout the time of the observations. From the time we start observing leg L2, feature 6 is the highest feature in it. The heights of features in L2 decrease as we go to the lowest one, which is feature 9.

\begin{figure}[!hbp]
\centering
\includegraphics[width=0.60\textwidth,clip=]{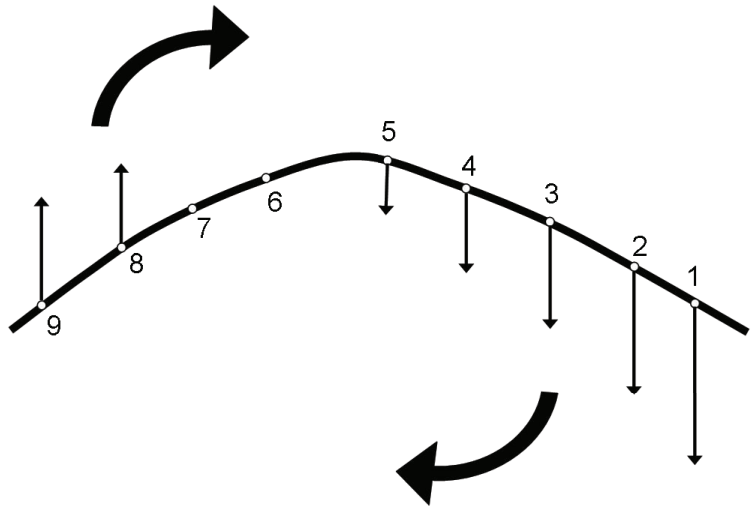}
\caption{A cartoon sketch, showing the prominence on 2010 April 13 projected against the solar disc, where, longitude is along the horizontal and latitude is along the vertical. The long curved line is the prominence, on which features 1 to 9 are marked as circles. The changes in latitude are shown as straight arrows, and the large curved arrows show the direction of prominence rotation, indicating a clockwise helical twist in the prominence spine.}\label{F:sk13apr}
\end{figure}

The middle panel of Figure~\ref{F:mb13apr} shows that the longitudes of the features (1 to 5) in L1 are significantly higher ($\sim20^{\circ}$ to $60^{\circ}$) than those in L2 (features 6 to 9) ($\sim0^{\circ}$ to $30^{\circ}$), i.e., leg L1 of the prominence is closer to the central meridian than leg L2. We can see that longitudes of all the features in L1 steadily increase throughout the period of observations, whereas longitudes of features in L2 remain almost constant before the eruption, but decrease appreciably once the eruption starts at 08:36 UT to reach an almost constant value of $\sim0^{\circ}$. The bottom panel of Figure~\ref{F:mb13apr} shows latitudes of the nine reconstructed features. Latitudes of features in leg L1 (triangles) are seen to decrease throughout the period of observations. Further we also observe that the decrease in latitude of the lowest feature (1) in L1 is the maximum, from $55^{\circ}$ to $43^{\circ}$. While the decrease for the highest feature (5) is the least, from $55^{\circ}$ to $50^{\circ}$. Contrary to leg L1, we find that latitudes of the two lowest features (8 and 9) in leg L2 show an increase of a few degrees, while latitudes of the two other features (6 and 7) do not show significant change. Figure~\ref{F:sk13apr} shows a cartoon sketch of the prominence projected against the solar disc. Changes in latitudes of the reconstructed features are shown as vertical arrows, while the overall direction of twist in the prominence is shown as thick curved arrows. Length of the vertical arrows are roughly indicative of the change in latitude for each feature.

Earlier studies have shown that prominences tend to travel in a non-radial equatorward direction during their eruption \citep{filippovea2001}. \citet{panasencoea2010} have studied such a deviation of prominences from their radial path in 3 events using stereoscopic reconstruction. Another dynamic form that a prominence exhibits during eruption is the writhe in its axis. \citet{gilbertea2001} have observed helical motion in prominences in He \textsc{I} 10830 \AA~images, while \citet{gilbertea2007} have tried to explain this motion by the means of a kinking filament. We propose that the prominence motion described in our study is due to superposition of these two separate motions. The first motion is the overall non-radial direction of propagation of the prominence, which directs the entire prominence towards the solar equator. While, the second motion is the helical twisting motion of the prominence spine. Leg L1 is at higher longitude than L2, hence L1 can be regarded as the western leg. Since changes in latitudes of features in L1 are quite large, as shown above, we believe that the twisting motion and non-radial motion are acting in the same direction for this leg (Figure~\ref{F:sk13apr}). On the other hand, for leg L2, which is at a lower longitude than L1, the changes in longitudes is small as compared to L1, hence we can say that the twisting motion and non-radial motion are acting in opposite directions for this leg. Therefore, the western leg, L1, shows a decrease in its latitude, while the eastern leg, L2, would have shown an increase in its latitude, but it is overpowered by the overall non-radial motion of the prominence. Thus we deduce that as the prominence erupts, its spine is seen to twist in a clockwise direction.

\begin{figure}[!hbp]
\centering
\includegraphics[width=0.60\textwidth,clip=]{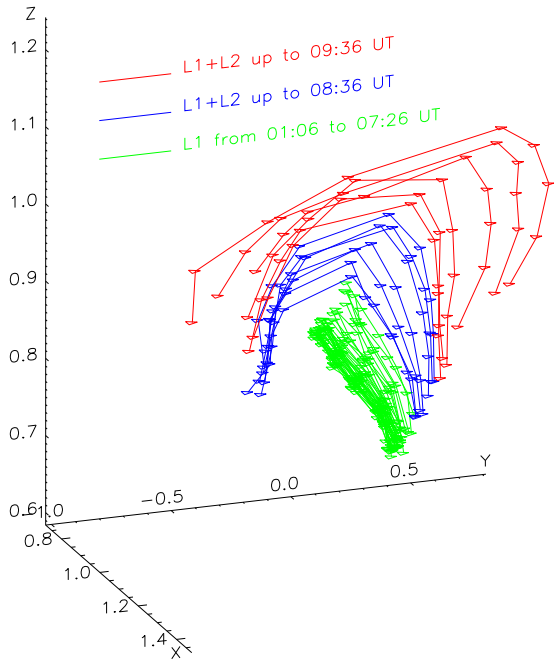}
\caption{Evolution of the erupting prominence on 2010 August 1 as seen in three dimensions in heliographic coordinate system. The position of the prominence determined by joining all the reconstructed points is shown at different instants of time, as marked on the plot.
%Cyan colour shows positions of the five features in leg L1 from 01:06 to 06:06 UT; green colour shows positions of the same features up to 07:26 UT; blue colour shows positions of all the nine features up to 08:36 UT; red colour shows positions of all the nine features up to 09:36 UT.
The coordinate system is centred on the Sun, with the Z axis along the solar rotation axis, and the X axis pointing towards the Earth. All the axes are in units of $\Rsun$.}\label{F:3D01aug}
\end{figure}

\begin{figure}[!hbp]
\centering
\includegraphics[width=0.60\textwidth,clip=]{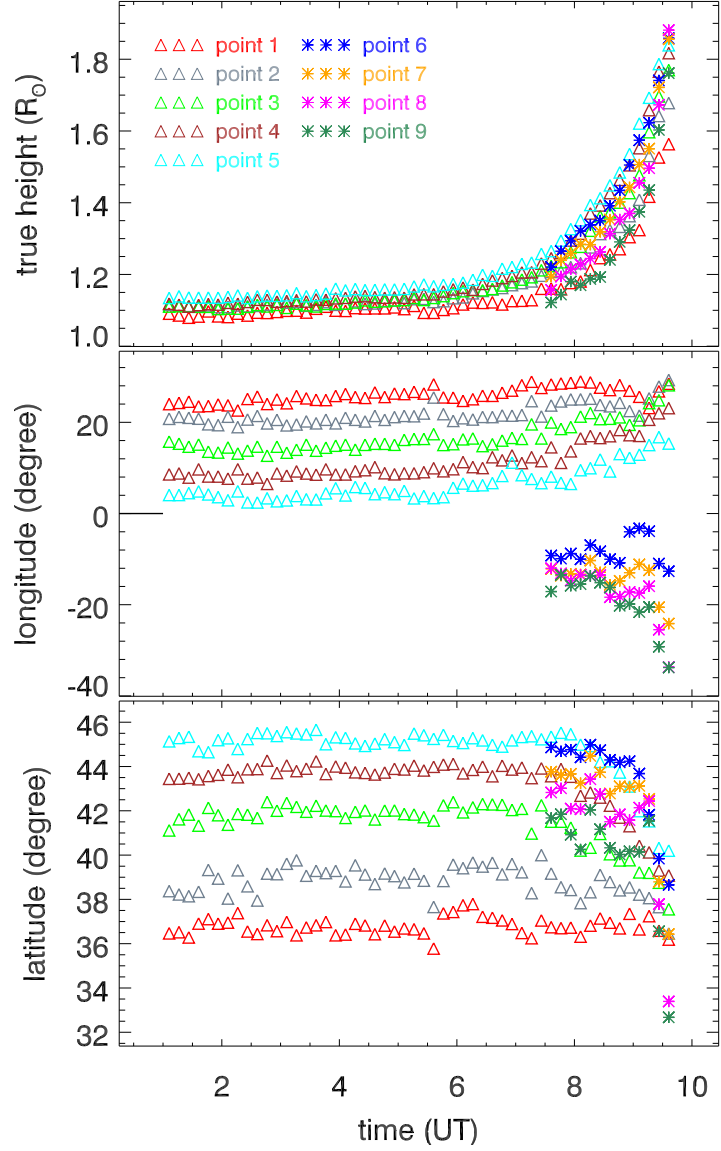}
\caption{The heliographic coordinates of different features of the prominence on 2010 August 1. Features 1 to 5 are from L1 to the top of spine, while features 6 to 9 are from spine to the lowest feature in leg L2 of the prominence.}\label{F:mb01aug}
\end{figure}

\subsection{2010 August 1 Prominence}

With respect to orientation relative to the two STEREO spacecraft (Figures~\ref{F:img13apr} and \ref{F:img01aug}), we can say that the two prominences in this study are mutually opposite. The prominence of 2010 August 1 initially appears like a hedgerow in 304~\AA~images from EUVI B, but only its spine is visible from EUVI A. The prominence leg seen on the right in EUVI B at \mbox{08:17 UT} in Figure~\ref{F:img01aug} is named L1, and the other leg is named as L2. For getting its shape in three dimension, we chose five features in leg L1 of the prominence. Leg L2 was obstructed from view by L1 for most of the time, and it became visible late into the eruption, from 07:36 UT onwards. Four more features along this leg were selected for reconstruction. As before, Figure~\ref{F:3D01aug} shows the prominence evolution in three dimensions in heliographic coordinates, and Figure~\ref{F:mb01aug} shows the true height along with the longitude and latitude of each of the nine features at different instants of time. Similar to Figure~\ref{F:3D13apr}, the numerous closely spaced lines in Figure~\ref{F:3D01aug} too indicate a slow rise for about six hours, before  a rapid eruption lasting around 2.5 hours. In Figure~\ref{F:mb01aug}, features 1 to 5 belong to L1, while features 6 to 9 belong to L2.

The top panel in Figure~\ref{F:mb01aug} gives true height of all features stereoscopically reconstructed in the prominence on 2010 August 1. The spread in heights of the features for this prominence is small compared to the spread seen for 2010 April 13 prominence. However, we still see that the height increases from the lowest feature (1), to the highest (5) in L1, and then decreases from the highest feature (6) to the lowest (9) in the L2, consistently throughout the observation period.

\begin{figure}[!hbp]
\centering
\includegraphics[width=0.60\textwidth,clip=]{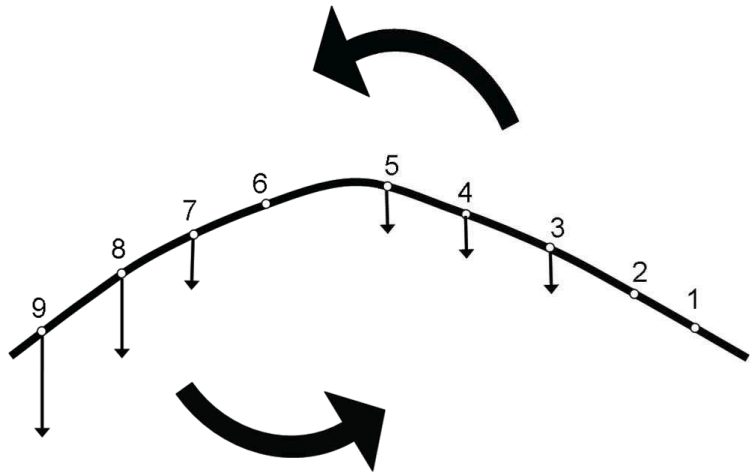}
\caption{A cartoon sketch, showing a top-down view of the prominence on 2010 August 1, where, longitude is along the horizontal and latitude is along the vertical. The long dark line is the prominence, while features 1 to 9 are marked as circles. The changes in latitude are shown as straight arrows, and the large curved arrows show the rotation direction of the prominence legs. This indicates an anticlockwise twist in the prominence spine.}\label{F:sk01aug}
\end{figure}

The latitudes and longitudes of the reconstructed features do not show as large a change as for those along the 2010 April 13 prominence. As we go from the lowest feature (1) in L1 through the spine to the lowest feature (9) in L2, we see that the longitude decreases more or less uniformly. Both longitude and latitude values for features in the L1 remain almost constant prior to the eruption. However, once the eruption starts at \mbox{07:06 UT}, features 3, 4 and 5, while lie close to the spine show an increase in longitude of about $10^{\circ}$ each, while the two lower features, 1 and 2, do not show any change in longitude. Leg L2 could be observed only once the eruption starts. With the exception of feature 6, longitudes of features 7, 8 and 9 are seen to decrease during the eruption. While features 7 and 8 decrease in longitude by the same amount of $\sim10^{\circ}$, the decrease in the lowermost feature in L2 is $15^{\circ}$. The latitudes of all the features except 1 and 2, are seen to decrease roughly by $6^{\circ}$.

An argument similar to the 2010 April 13 event about the two separate motions experienced by the prominence can be extended towards this event. Leg L1 is at higher longitude that L2, hence L1 can be termed as the western leg, and L2 as the eastern leg. As explained above, features 1 and 2 do not show much change either in their latitudes or longitudes. Features 3, 4 and 5 in the western leg however, show a small decrease in latitude, and a small increase in longitude. While, features in the eastern leg (L2) show a significant change in both latitude and longitude. In this case we find that it is the eastern leg that experiences a stronger decrease in latitude compared to the western leg, because of the opposing motions of twist and non-radial equatorward direction of propagation of the prominence during its eruption. Hence, we can infer that the prominence on 2010 August 1 showed a twist in its axis in the anticlockwise direction. The prominence projected against the solar disc is shown in the form of a cartoon sketch in Figure~\ref{F:sk01aug}. Changes in latitudes of the reconstructed features are shown as vertical arrows, while the overall direction of twist in the prominence is shown as thick curved arrows. Length of the vertical arrows are roughly indicative of the change in latitude for each feature.

\subsection{Acceleration of the Prominences}

Several studies have found that a prominence is observed to display two phases during their eruption, the slow rise and the fast eruption. \citet{sterlingmoore2005} have found a slow rise phase with a constant velocity, and then an accelerating fast eruptive phase. On the other hand \citet{joshisrivastava2007} have reported a small acceleration even in the slow rise phase, while \citeauthor{sterlingmoore2004a} (\citeyear{sterlingmoore2004a,sterlingmoore2004b}) have found the prominence to rise with constant velocities during both the phases. It may be noted that these conclusions were based on observations from a single viewpoint. For the two prominences studies here, we identified a time of eruption after which the true height starts increasing very rapidly. A 2$^{\textrm{nd}}$ order polynomial was fitted separately to the height for time before and after the time of eruption. The fitted function was used to obtain the speed and acceleration of all the features of the prominences.

\subsubsection{The slow rise phase}

The two prominences analysed by us exhibit slow rise followed by a fast eruption. From Figure~\ref{F:mb01aug} top panel, we see that the prominence on 2010 April 13 increased in height over a period of almost five hours before its eventual eruption which commenced at 08:36 UT. The features in leg L1 show an almost uniform motion during this phase, rising with an average acceleration of $67~\rm{cm~s^{-2}}$. However, features in L2 showed a wide range of acceleration, ranging from $-46~\rm{cm~s^{-2}}$ for feature 9 to $123~\rm{cm~s^{-2}}$ for feature 7. We believe that the large spread in acceleration values for leg L2 is because of the lesser number of points available in the slow rise phase for fitting a 2nd order polynomial.

A similar procedure was employed to obtain acceleration of the features for 2010 August 1 prominence. The eruption for this event was found to begin at 07:06 UT. We once again find an almost constant acceleration in leg L1; its average value being $31~\rm{cm~s^{-2}}$ for features 2 to 5. Feature 1, however, showed a relatively low acceleration at $4~\rm{cm~s^{-2}}$. Since leg L2 could be observed only once the eruption started, it was not possible to observe the slow rise in this leg prior to the eruption.

The values for constant acceleration of the features during the slow rise of prominences obtained here are considerably higher than those obtained by \citet{joshisrivastava2007} which were in the range $4-12~\rm{cm~s^{-2}}$.

\subsubsection{The fast eruptive phase}

On fitting the true heights of all the features in the eruptive phase for both the prominences with a polynomial function, it was found that they rose with a constant acceleration. For the 2010 April 13 prominence, each of the five features in leg L1 showed a constant value of acceleration ranging from $9~\rm{m~s^{-2}}$ to $12~\rm{m~s^{-2}}$, with an average value of $11~\rm{m~s^{-2}}$. Whereas, the four features in leg L2 showed acceleration in the range $2~\rm{m~s^{-2}}$ to $8~\rm{m~s^{-2}}$, with an average value of $5~\rm{m~s^{-2}}$. The significant difference between values of average acceleration in the two legs during the eruptive phase can be attributed to the two forces acting on the prominences. This prominence is shown to twist in clockwise direction, which means that in the western leg, L1, the two forces, viz, helical twist and non-radial motion, are acting in the same direction, but they are acting in opposite directions on the eastern leg, L2. Therefore, L1 shows a higher value of average acceleration in the eruptive phase, whereas, L2 shows a relatively lower value of average acceleration.

For the 2010 August 1 prominence too, a constant acceleration for each reconstructed feature was derived. For the five features in L1, the values range from $9~\rm{m~s^{-2}}$ to $11~\rm{m~s^{-2}}$, with an average value of $10~\rm{m~s^{-2}}$, while for the four features in L2 had their acceleration values ranging from $16~\rm{m~s^{-2}}$ to $23~\rm{m~s^{-2}}$ with an average of $20~\rm{m~s^{-2}}$. A very similar argument as put forward above can be used to explain the markedly distinct values of average acceleration observed in the two legs of this prominence as well. As shown earlier, this prominence twists in the anticlockwise direction during eruption. Thus, in the eastern leg, L2, the twisting motion and the non-radial propagation act in the same direction giving rise to a higher acceleration, compared to the western leg, L1, wherein these two motions act in mutually opposite directions.

We find that the values for constant acceleration of the features during the fast eruptive phase of prominences are slightly lower than the maximum acceleration in the range of $3-77~\rm{m~s^{-2}}$ obtained by \citet{joshisrivastava2007}, but an order of magnitude lower than the value found by \citet{sterlingmoore2005}, which is $1.0~\rm{m~s^{-2}}$. This may be due to the fact that the prominences analysed by \citet{joshisrivastava2007} and the two prominences in the present study are quiescent in nature, whereas those analysed by \citet{sterlingmoore2005} are active region prominences.

%%%%%%%%%%%%%%%%%%%%%%%%%%%%%%%%%%%%%%%%%%%%%%%%%%%%%%%%%%%%%%%%%%%%%%%%%
%%% New section
%%%%%%%%%%%%%%%%%%%%%%%%%%%%%%%%%%%%%%%%%%%%%%%%%%%%%%%%%%%%%%%%%%%%%%%%%
\section{Conclusions}\label{S:conclu}

Three-dimensional reconstruction of two northern-hemisphere polar crown eruptive prominences on 2010 April 13 and 2010 August 1 was carried out using 304~\AA~images from EUVI instrument on board the twin STEREO spacecraft. For this purpose a stereoscopic reconstruction technique developed by us was used. Both the prominences had both their legs anchored to the photosphere during major part of the eruptive phase. Several features along each leg of the prominence were chosen and carefully followed in each pair of images to obtain the true coordinates, and hence the true shape of the prominences at as many instants of time as possible.

The variations in true longitude and latitude of the reconstructed features in both the legs of the prominences was observed as they erupted. We found the variation to be because of an interplay of two motions: the overall non-radial equatorward motion of the prominence towards the equator, and the helical twist in the prominence spine. These variations in latitude and longitude of the reconstructed features suggest that the spine of the prominence on 2010 April 13 twisted in a clockwise direction while spine of the prominence on 2010 August 1 twisted in an anticlockwise direction during eruption.

Our three-dimensional study of prominence kinematics showed two distinct phases of eruption: the slow rise and the fast eruptive phase, as shown by previous studies by \citet{sterlingmoore2005} and \citet{joshisrivastava2007} which were based on projected plane-of-sky observations. The acceleration determined is different at different features along the prominences, but it is constant if just one feature is considered. The acceleration values in the fast eruptive phase show strong grouping in each leg of the prominence in both the events analysed. The net effect of the two motions, viz, non-radial propagation and helical twist in spine, produce a higher average acceleration in the western leg, $11~\rm{m~s^{-2}}$, compared to the eastern leg, $5~\rm{m~s^{-2}}$, of the prominence on 2010 April 13. While, for 2010 August 1, these two forces act to give rise to higher average acceleration in eastern leg, $20~\rm{m~s^{-2}}$, compared to the western leg, $10~\rm{m~s^{-2}}$.

The study of acceleration in the prominence legs as a response to the two dynamic motions experienced by them should be considered for a better understanding of prominence eruptions in future studies.

The authors thank the STEREO/SECCHI consortium for providing the data. The SECCHI data used here were produced by an international consortium of the Naval Research Laboratory (USA), Lockheed Martin Solar and Astrophysics Lab (USA), NASA Goddard Space Flight Center (USA), Rutherford Appleton Laboratory (UK), University of Birmingham (UK), Max-Planck-Institut for Solar System Research (Germany), Centre Spatiale de Li$\grave{\textrm e}$ge (Belgium), Institut d'Optique Theorique et Appliqu$\acute{\textrm e}$e (France), Institut d'Astrophysique Spatiale (France). The authors are very grateful to Bernd Inhester for providing valuable suggestions on the reconstruction technique developed in this study. Work by N.S. partially        contributes to the research on collaborative NSF grant \mbox{ATM-0837915} to Helio Research.

%%%%%%%%%%%%%%%%%%%%%%%%%%%%%%%%%%%%%%%%%%%%%%%%%%%%%%%%%%%%%%%%%%%%%%%%%
%%% New section
%%%%%%%%%%%%%%%%%%%%%%%%%%%%%%%%%%%%%%%%%%%%%%%%%%%%%%%%%%%%%%%%%%%%%%%%%
\section*{Appendix}\label{S:apndx}

\begin{figure}[!hbp]
\centering
\includegraphics[width=0.75\textwidth]{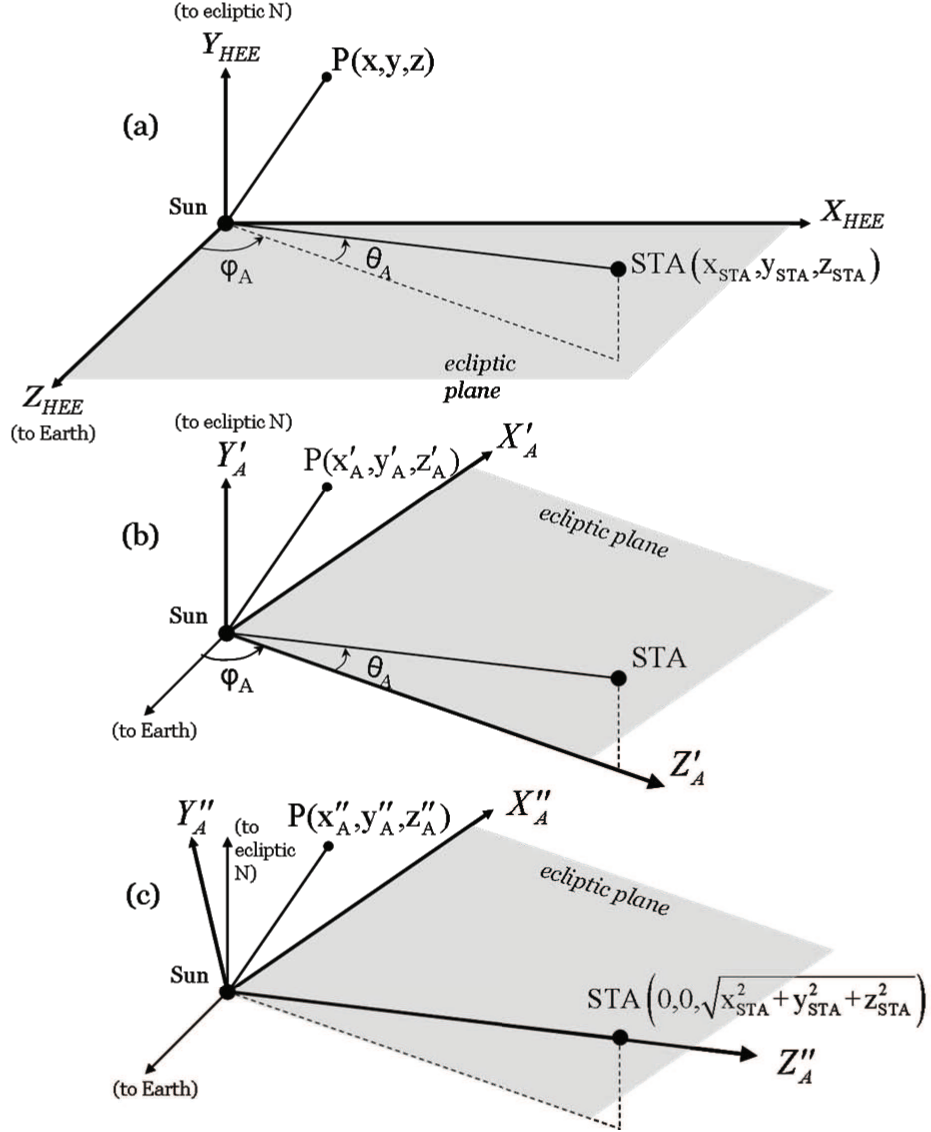}
\caption{(a) Shows a point $P(x,y,z)$ in rHEE system. $\phiA$ and $\thtA$ are longitude and latitude of STA. (b) rHEE system is rotated in anti-clockwise direction about $Y_{HEE}$ through angle $|\phiA|$. (c) Rotation of the $X'_{A}Y'_{A}Z'_{A}$ coordinate system about the $X'_{A}$ axis through angle $-\thtA$, where the sign of $\thtA$ takes care of the sense of rotation.}\label{F:STArot}
\end{figure}

We will determine true position of a point $P(x,y,z)$ in the HEE coordinate system and then convert it to the more familiar heliographic system. The HEE coordinate system is centred at the origin, wherein, its z-axis points towards the ecliptic north pole, x-axis is the Sun-Earth line, while y-axis completes the right-handed triad \citep{hapgood1992,thompson2006}. However, for the sake of convenience, we have changed labels of the axes of the HEE coordinate system. Therefore, the y-axis ($Y_{HEE}$) points towards the ecliptic north pole, z-axis ($Z_{HEE}$) is the Sun-Earth line, and x-axis ($X_{HEE}$) completes the right-handed triad (Figure~\ref{F:STArot}). To maintain a distinction between the real HEE system and the newly defined HEE system with its axes relabelled, we name the latter as rHEE.

We carry out the reconstruction by first rotating the rHEE coordinate system in such a manner that the $x$ and $y$ coordinates in the rotated system are same as the $x$ and $y$ coordinates of the image as seen by the spacecraft STEREO A (STA) and B (STB). Under the assumption of affine geometry, the observable angular separation of a solar feature can be converted to physical distance by taking into account the platescale of the image. Since we know coordinates of the two spacecraft in rHEE system, we can execute the rotations by making use of the rotation matrices. These transformation equations are then used to obtain the true coordinates $(x,y,z)$ of point $P$ in HEE system.

First consider STA alone. Let $\phiA$ and $\thtA$ be its longitude and latitude in rHEE system, as shown Figure~\ref{F:STArot}a. We need to perform two rotations on the rHEE system to orient it such that coordinates of $P$ in the rotated system, $x''_{A}$ and $y''_{A}$, are same as its $x$ and $y$ coordinates in the image in STA. The two rotations are:
\begin{enumerate}
\item Longitude of A, $\phiA$, is always considered to be positive. Hence, to align the $Z_{HEE}$ axis with the $Y_{HEE}$-Sun-STA plane, we need to rotate the rHEE coordinate system about the $Y_{HEE}$ axis through angle $\phiA$ (Figure~\ref{F:STArot}b), to give the modified axes $X'_{A}, Y'_{A}~\rm{and}~Z'_{A}$ (Equation~\ref{Q:STAr1}).

\begin{equation}
\begin{pmatrix} x'_{A}\\ y'_{A}\\ z'_{A} \end{pmatrix} = 
\begin{pmatrix} 
	\cos{\phiA}&0&-\sin{\phiA}\\ 0&1&0\\ \sin{\phiA}&0&\cos{\phiA}
\end{pmatrix}
\begin{pmatrix} x\\ y\\ z \end{pmatrix}
\label{Q:STAr1}
\end{equation}

\item Latitude of A, $\thtA$ varies periodically between $\pm 0^{\circ}.13$. Hence, to align the $Z'_{A}$ axis with STA line-of-sight, we need to rotate the $X'_{A}Y'_{A}Z'_{A}$ coordinate system about the $X'_{A}$ axis through angle $-\thtA$ (Figure~\ref{F:STArot}c), to give the modified axes $X''_{A}, Y''_{A}~\rm{and}~Z''_{A}$ (Equation~\ref{Q:STAr2}).

\begin{equation}
\begin{pmatrix} x''_{A}\\ y''_{A}\\ z''_{A} \end{pmatrix} = 
\begin{pmatrix}
	1&0&0\\ 0&\cos{\thtA}&-\sin{\thtA}\\ 0&\sin{\thtA}&\cos{\thtA} 
\end{pmatrix}
\begin{pmatrix} x'_{A}\\ y'_{A}\\ z'_{A} \end{pmatrix}
\label{Q:STAr2}
\end{equation}

\end{enumerate}
We thus have STA looking down the $Z''_{A}$ axis. In other words, the $X''_{A}-Y''_{A}$ plane becomes the plane of the sky as seen from STA. The two rotations are illustrated with the help of Fig~\ref{F:STArot}.
Combining Equations~\ref{Q:STAr1} and \ref{Q:STAr2}, gives us:
\begin{equation}
\begin{pmatrix} x''_{A}\\ y''_{A}\\ z''_{A} \end{pmatrix} = 
\begin{pmatrix}
	x\cos{\phiA} - z\sin{\phiA}\\
	- x\sin{\phiA}\sin{\thtA} + y\cos{\thtA} - z\cos{\phiA}\sin{\thtA}\\
	x\sin{\phiA}\cos{\thtA} + y\sin{\thtA} + z\cos{\phiA}\cos{\thtA}\\	
\end{pmatrix}
\label{Q:STArc}
\end{equation}

On applying identical transformations to STEREO B, we get the following equation, 

\begin{equation}
\begin{pmatrix} x''_{B}\\ y''_{B}\\ z''_{B} \end{pmatrix} = 
\begin{pmatrix}
	x\cos{\phiB} - z\sin{\phiB}\\
	-x\sin{\phiB}\sin{\thtB} + y\cos{\thtB} - z\cos{\phiB}\sin{\thtB}\\
	x\sin{\phiB}\cos{\thtB} + y\sin{\thtB} + z\cos{\phiB}\cos{\thtB}\\	
\end{pmatrix}	
\label{Q:STBrc}
\end{equation}
where, $(x''_{B},y''_{B})$ are the image coordinates as seen from STB, and also coordinates of point $P$ in the twice-rotated rHEE system.

\sloppy We thus have four equations, two each from Equations~\ref{Q:STArc} and \ref{Q:STBrc}, with four variables $(x''_{A},y''_{A},x''_{B},y''_{B})$ and 3 unknowns $(x,y,z)$ to be determined. In order to solve this over-determined system, we invoke the epipolar constraint. The two STEREO spacecraft along with the point to be reconstructed form a plane. This plane is known as an epipolar plane \citep{inhester2006}. Since by definition, each of the two spacecraft lie on all the epipolar planes, the projection of this plane onto the image from any of the two spacecraft is a straight line. The epipolar constraint requires that any feature lying on a certain epipolar plane as seen through one spacecraft, must lie on the same epipolar plane as seen through the other spacecraft too.

Once we identify a feature either in image from STA or from STB, it is possible to determine projection of the epipolar plane passing through the feature selected, on the image from the second spacecraft. For this, the two image coordinates from the first spacecraft along with an assumed pair of points are converted to the observer-independent heliocentric Earth equatorial (HEEQ) coordinate system. The HEEQ coordinates of these points are then converted to the image coordinates of the second spacecraft to obtain the projection of the two points on its image \citep{thompsonwei2010}. The projection forms a line passing through the feature selected in the first image. The images can be oriented so that the STEREO mission plane (the plane containing STA and STB and passing through the Sun's centre) is along the horizontal in both the images (\citeauthor{inhester2006}, \citeyear{inhester2006} \citeyear{inhester2009}). As a result of this orientation, any epipolar plane projects as a horizontal line in the two images. This line then passes through the selected feature in the second image, thus constraining the value of the y-coordinate of the feature in the second image to a known value, which leaves us with three equations.

Suppose, we select the feature in image from STA first, then we know $x''_{A}, y''_{A}$ and $x''_{B}$. The value of $y''_{B}$ is known because of the horizontal epipolar line passing through the feature in the image from STB. Thus, we get Equation~\ref{Q:xyz} given in the text. However, if we select the feature in image B first, then only the equation for $y$ changes as follows:

\begin{equation}
y\,=\,\frac{y''_{B}}{\cos{\thtB}}\, + \,\Big(\frac{x''_{B}}{\tan(\phiA-\phiB)} - \frac{x''_{A}}{\sin{(\phiA-\phiB)}}\Big)\tan{\thtB}
\label{Q:y}
\end{equation}
If we consider a human error of 3 pixels made while selecting a feature in the EUVI image, we find the errors in distance, longitude and latitude are $0.02~\Rsun$, $2^{\circ}$ and $0^{\circ}.5$ respectively. Since, we can apply this technique equally well to coronagraph images, we find that an error of 3 pixels made while selecting a feature in the COR1 image, the error made in determining the height of the feature comes out to be $0.12~\Rsun$, while the errors in longitude and latitude are the same as obtained for EUVI images.

%%%%%%%%%%%%%%%%%%%%%%%%%%%%%%%%%%%%%%%%%%%%%%%%%%%%%%%%%%%%%%%%%%%%%%%%%
%%% Bibliography
%%%%%%%%%%%%%%%%%%%%%%%%%%%%%%%%%%%%%%%%%%%%%%%%%%%%%%%%%%%%%%%%%%%%%%%%%

%%\bibliographystyle{apj}
%%\bibliography{prom_twist}

\end{document}